\documentclass[a4paper,fleqn,usenatbib]{mnras}

\usepackage[T1]{fontenc}
\usepackage{ae,aecompl}

\usepackage{graphicx}	
\usepackage{amsmath}	
\usepackage{amssymb}	
\usepackage{threeparttable}
\usepackage{pdflscape}

\title[Local RR Lyrae]{Local RR Lyrae stars: native and alien} 

\author[R. Zinn et al.]{
R. Zinn,$^{1}$\thanks{E-mail: robert.zinn@yale.edu}
X. Chen,$^{2}$
A. C. Layden$^{2}$
and D. I. Casetti-Dinescu$^{3}$
\\
$^{1}$Department of Astronomy, Yale University, P.O. Box 208101, New Haven, CT 06520-8101, USA\\
$^{2}$Physics \& Astronomy Department, Bowling Green State University, Bowling Green, OH 43403, USA\\
$^{3}$Department of Physics, Southern Connecticut State University, 501 Crescent Street, New Haven, CT 06515, USA
}

\date{Accepted XXX. Received YYY; in original form ZZZ}

\pubyear{2019}

\begin{document}
\label{firstpage}
\pagerange{\pageref{firstpage}--\pageref{lastpage}}
\maketitle

\begin{abstract}
Measurements of [Fe/H] and radial velocity are presented for 89 RR Lyrae (RRL) candidates within 6 kpc of the Sun.  After the removal of two suspected non-RRL, these stars were added to an existing database, which yielded 464 RRL with [Fe/H] on a homogeneous scale.  Using data from the \textit{Gaia} satellite (Data Release 2), we calculated the positions and space velocities for this sample.  These data confirm the existence of a thin-disc of RRL with [$\alpha$/Fe]$\sim$ solar.  The majority of the halo RRL with large total energies have near zero angular momenta about the Z axis.  Kinematically these stars closely resemble the Gaia-Sausage/Gaia-Enceladus stars that others have proposed are debris from the merger of a large galaxy with the Milky Way.  The metallicity and period distributions of the RRL and their positions in the period-amplitude diagram suggest that this disrupted galaxy was as massive as the Large Magellanic Cloud and possibly greater.
\end{abstract}




\section{Introduction}

Investigations of the structure of the Milky Way (MW) often employ one type of star or star cluster and use their distributions in space, kinematics, chemical composition, and age to probe the MW's current structures, with the goal of revealing their origins.  Each probe has its advantages and limitations in terms of their identification, number density, luminosity distribution, and the ease of measurement of the fundamental data.  Time-honored examples of probes are the subdwarfs and other main-sequence stars, globular and open clusters, red giants, classical Cepheid variables, and RR Lyrae (RRL) variables.  Because RRL are easily identified by their light variations and have nearly uniform, well-determined, and large mean luminosities ($\sim 50 L_{\odot)}$), they have been used for decades to trace old stellar populations.  Following the example of many previous observers \citep[e.g.,][]{layden95a,kepley07,morrison09,kinman12,dambis13,marsakov18a}, we use an enlarged sample of well-studied RRL near the Sun to investigate the halo and disc populations.  Our motivation stems from many recent theoretical and observational investigations of the origins of these populations. 

Simulations of the formation of MW-size spiral galaxies in the framework of modern cosmology predict that their stellar halos have two components \citep[e.g.,][]{zolotov09,tissera12,pillepich15,cooper15,monachesi19}.  One, which is frequently called the accreted halo, is made up of stars that formed in dwarf galaxies that were accreted and ultimately destroyed by merging with the large galaxy early in its evolution.  The second component, which is commonly called the `in situ' halo, is made up of stars that formed in the inner disk and bulge of the large galaxy.  Merger events later dislodged these stars from their birth sites and placed them on halo orbits.  Because of this orbit modification, the in situ halo has been also called the `kicked out' component of the halo by some authors.  The in situ halo of the simulations should not be confused with the halo model of the classic paper by \citet[][ELS]{eggen62} in which the halo stars were the first stars to form in a radially collapsing gas cloud and owe their eccentric orbits to the cloud's rapidly changing gravitational potential.  The modern simulations do not identify a halo component that closely resembles the ELS model.  The in situ halo stars are predicted to be more metal rich and have more prograde rotation than the majority of the accreted halo stars because of their origin in the nascent disk. The in situ halo and the thick disc are predicted to be related in the sense that the merger events that produced the in-situ halo may have also contributed stars to the thick disk, either directly from the disrupted galaxies or by heating the thin disc. 

Observations over the past 40 years have shown that the MW halo has a sizable accreted component in agreement with the simulations \citep{sz78,norris86,carney90,majewski93,chiba00,helmi99,mackey04,belokurov07c,bell10,carollo10,schuster12}.  Evidence for a 'dual halo' has been growing in the last decade, whose components are often identified with the accreted and in situ halos of the simulations \citep[e.g.,][]{carollo10, nissen10, bonaca17}, but the size of the in situ halo has been debated \citep{hawkins15,deason17}. In the past year, several studies \citep[][and refs. therein]{belokurov18b,haywood18,helmi18,koppelman18,kruijssen19,myeong19,donlon19} have provided strong evidence that the MW has experienced one, and perhaps more than one, major merger roughly 10 Gyrs ago that was of sufficient in size to substantially heat any pre-exisiting disc.  The origin of the thick disk may now be understood, but there remains debate about whether a separate in situ halo has been uncovered (e.g.,\citep{haywood18,mackereth19a,dimatteo19,gallart19}). 
    
In this paper, we report the results of a spectroscopic survey of RRL stars within 6 kpc of the Sun.  In several respects, it resembles the earlier survey by \citet{layden94,layden95a}, which remains a major source of data on the nearby RRL.   Ab initio, the present survey was motivated by the first observational results for a dual halo and the first simulations showing the accreted and in situ halos.  The availability a new catalogue of bright RRL \citep{Pojmanski02} and a new catalogue of proper motions for southern stars \citep{girard11} made feasible the goal of measuring the space velocities and metallicities of hundreds of RRL following the methodology of \citet[][L94]{layden94}.   The decommissioning of the spectrograph that was used for this survey not long after it started severely limited the data that was obtained, and here we report the results of observing only 89 stars, many of which do not have previous measurements of [Fe/H].  After reporting our measurements, we add this new sample to an existing one, and discuss the properties of 464 nearby RRL.

\section{Observations and Reductions}    \label{sec_obs}

\subsection{Spectroscopy}    \label{ssec_spm}

In planning this study, a sample of about one hundred ab-type RRL was selected from the ASAS online
catalog\footnote{Visit http://www.astrouw.edu.pl/asas/.}. 
Queue-scheduled observations were requested on the 1.5-m telescope and
Cassegrain Spectrograph located on Cerro Tololo in Chile and operated
by the Small and Moderate Aperture Research Telescope System (SMARTS)
Consortium.  The Loral 1K CCD (having 15 $\micron$ pixels in a 1200 by
800 array) 
was used with grating 26/Ia at a tilt of 16.14$\degr$,
providing spectra covering 3650 to 5400 \AA\ at 1.47 \AA\ pix$^{-1}$
resolution.  A slit width of 0.11 mm (2.0 arcsec) was used
independent of seeing to facilitate the queue-scheduled observations.

For most observations, HeAr comparison spectra were taken immediately
before and after a set of three stellar exposures (30 to 240 s each)
in order to monitor and remove the effects of instrumental flexure.
In total, 171 such ``visits'' were made to 89 separate RRL stars
between 2009 August and 2011 February, yielding 525 useable spectra.
In addition we obtained 75 spectra in 25 visits to six non-variable
stars from Table 4 of L94 to serve as equivalent width standard stars.
 
We performed the usual CCD image processing steps of bias subtraction
and flat-field correction using IRAF\footnote{The Image Reduction and
Analysis Facility (IRAF) was distributed by the National Optical
Astronomy Observatory, which is operated by the Association of
Universities for Research in Astronomy (AURA) under a cooperative
agreement with the National Science Foundation.}, then extracted each
two-dimensional spectrum by summing across the stellar spectrum and
subtracting the sky background using apertures on either side of the
stellar spectrum.  An extraction was made of each comparison lamp
exposure in a visit using the identical parameters used for each
stellar spectrum, and the two lamp spectra (one preceding and one
following) were used to provide a wavelength calibration for each
stellar spectrum.  This involved spline curve fits to the the pixel
locations of 47 HeAr emission lines of known wavelength in each
comparison spectrum, yielding a typical RMS scatter of 0.19 \AA\ or 13
km\ sec$^{-1}$ in radial velocity.  Finally, we continuum-normalized
each wavelength-calibrated stellar spectrum by fitting and dividing by
a spline curve to remove the wavelength-dependent instrumental
response.

Table~\ref{tab_coords} lists, by their ASAS catalog names, the 89 RRL
stars observed.
The second and third
columns of Table~\ref{tab_coords} list the equatorial coordinates, in
degrees at epoch J2000.0, from the Yale/San Juan Southern Proper
Motion Program (SPM4) \citep{girard11}.  The positional uncertainties
are typically 23~mas, much more precise than the ASAS coordinates (the
two sets of coordinates typically differ by 5~arcsec).  Four stars were not 
found in the SMP4 (UZ~Eri, LR~Eri, XX~Pup, and 040312-1951.2), so we show the ASAS equatorial coordinates for these stars in Table~\ref{tab_coords}.
We used the equatorial coordinates of each star to find the star's Galactic coordinates ($l$, $b$), shown in columns four and five of the table.  
Most of the stars had $230^\circ < l < 360^\circ$ and $-90^\circ < b < +10^\circ$.
One of the stars, 083412-6836.1, is no longer classified as a RRL, but instead as a CWB variable (BL Her or Anomalous Cepheid) by the recent ASAS-SN survey \citep{jayasinghe19}.  We include it in our analysis because it resembles a RRab spectroscopically, although we caution that the value of [Fe/H] that we obtain may be systematically in error. This star is excluded from our discussion of the kinematics of the local RRL.

\begin{table*}
\caption{Coordinates and Reddening}
\begin{threeparttable}
\label{tab_coords}
\begin{tabular}{cccccccc}
\hline
ASAS~ID & RA (J2000) & Dec (J2000) & $l$ & $b$  & $E(B-V)$ & Alt.~ID & Note\\
\hline
000602-3654.3 &   1.5081394 & --36.9041704 & 344.58 & --76.30 & 0.008 & WW~Scl    & ... \\
000621-3517.2 &   1.5868393 & --35.2868640 & 349.87 & --77.39 & 0.011 & CD~Scl & ... \\
001543-5853.1 &   3.9256951 & --58.8835238 & 311.55 & --57.64 & 0.010 & UZ~Tuc    & ... \\
002418-7616.9 &   6.0662235 & --76.2819659 & 305.05 & --40.73 & 0.059 & T~Hyi     & ... \\
002443-6949.7 &   6.1758203 & --69.8288655 & 306.31 & --47.12 & 0.022 & ...   	  & ... \\
002843-4400.4 &   7.1795756 & --44.0062839 & 316.64 & --72.51 & 0.008 & NSV~00174 & ... \\
...           & ...         & ...          & ...    & ...     &  ...  & ...   	  & ... \\
221843-5652.4 & 334.6765399 & --56.8733809 & 334.52 & --49.84 & 0.015 & ...   	  & ... \\
231648-4539.0 & 349.1999451 & --45.6486414 & 342.09 & --63.63 & 0.007 & ...   	  & ... \\
232246-4641.5 & 350.6939864 & --46.6922994 & 338.80 & --63.79 & 0.005 & NSV~14530 & ... \\
\hline 
\end{tabular}
\begin{tablenotes}[normal,flushleft]
\item Table \ref{tab_coords} is published in its entirety in the online edition; a portion is shown here for guidance regarding its form and content.
\end{tablenotes}
\end{threeparttable}
\end{table*}

We searched the Infrared Science Archive (IRSA)
database\footnote{Visit \url{http://irsa.ipac.caltech.edu/applications/DUST/}.} at these coordinates to retrieve the interstellar redding value, $E(B-V)$, from the \citet{schlafly11} recalibration of the COBE dust maps \citep{schlegel98}.
Three of our targets were in front of the Large Magellanic Cloud and
had anomalously large reddenings.  For these stars (045129-7137.7, 045426-6626.2, and VW~Dor), we replaced the IRSA reddening value with the mean value
from eight surrounding RRL: $E(B-V) = 0.049$ mag (standard deviation
0.023 mag).  The resulting values are shown in column six of Table~\ref{tab_coords}.  For some targets, we include a cross-referenced alternate identification in the seventh column of Table~\ref{tab_coords}.  The final column contains notes about selected stars.

\subsection{ASAS Photometry and Period Revision}     \label{ssec_asas}

The ASAS project \citep{Pojmanski02} used a $2048 \times 2048$ pixel
CCD behind a 200 mm, $f/2.8$ camera lens to obtain time-series,
wide-field, low-resolution, $V$-band images across most of the
Southern sky.  We downloaded from the ASAS website the light
curve chacteristics (magnitude at maximum brightness, amplitude of
variation, period, epoch of maximum light, and variable type) output
by the automated ASAS analysis pipeline.

Inspection of several light curves suggested that improvements could
be made to the period, and hence other derived light curve properties.
We therefore downloaded the photometric data for each star (magnitude
vs. time) from the ASAS website.  For consistency, we used the
magnitude from aperture \#2 for each observation.  We rejected photometry 
characterized by grades other than `A' or `B', leaving many hundreds
of individual observations per star.

Then, we used the Phase Dispersion Minimization  
program \citep{stellingwerf78} implemented in IRAF to seek the best
period for each star. For about half of the 89 stars,  our revised period resulted in a phased light curve with less scatter than in the original ASAS result. 
We also estimated the uncertainty in each period, $\epsilon_P$, by
finding the period offset that resulted in a light curve with
noticeably increased scatter.

We then fit the phased light curve of each star with a series of
templates using the procedure described in \citet{layden98}.  This
resulted in a number of useful properties including the intensity-mean
magnitude, $\langle V \rangle$, the maximum and minimum brightness
(all calculated from the best-fitting template, $T$; see Fig.~1 of
\citet{layden98}) and the RMS scatter of the observed points around
that template, along with the epoch of maximum light, $E_{max}$.
These values are shown in Table~\ref{tab_asas}, along with the period,
$P$, its uncertainty, and the number of photometric observations,
$N_{obs}$.

\begin{table*}
\caption{ASAS Photometry}
\begin{threeparttable}
\label{tab_asas}
\begin{tabular}{cccrccccccc}
\hline
ASAS~ID &  Alt.~ID  &  Period &     $\epsilon_P ~^a$     & $\langle V \rangle$    & $V_{max}$  &
$V_{min}$     & $RMS$      & $T^b$     & $E_{max}$     & $N_{obs}$   \\
\hline
000602-3654.3 & WW~Scl    & 0.784880 &  2 & 13.62 & 13.22 & 13.93 & 0.20 & 6 & 2453512.43518 &  383 \\
000621-3517.2 &  CD~Scl & 0.577670 & 20 & 13.54 & 12.91 & 14.01 & 0.21 & 2 & 2452985.59114 &  387 \\
001543-5853.1 &  UZ~Tuc & 0.625311 &  5 & 13.17 & 12.51 & 13.56 & 0.15 & 1 & 2454736.14325 &  724 \\
002418-7616.9 & T~Hyi   & 0.568740 & 20 & 13.70 & 13.14 & 14.17 & 0.28 & 6 & 2452226.51785 &  968 \\
002443-6949.7 &  ... & 0.727967 & 10 & 11.11 & 10.93 & 11.29 & 0.06 & 7 & 2454367.29273 &  585 \\
002843-4400.4 & NSV~00174  &0.599972 & 10 & 13.00 & 12.54 & 13.35 & 0.14 & 4 & 2452206.22194 &  492 \\
...     &      ... & ...     &      ... & ... &  ...  &  ...  &  ...  & ...  & ...  \\
221843-5652.4 & ...  & 0.721518 & 10 & 13.76 & 13.38 & 14.06 & 0.21 & 6 & 2453543.34235 &  400 \\
231648-4539.0 &  ... & 0.894911 & 20 & 13.16 & 12.95 & 13.32 & 0.17 & 6 & 2453749.12942 &  503 \\
232246-4641.5 & NSV~14530  & 0.553218 &  2 & 12.54 & 11.98 & 12.94 & 0.08 & 2 & 2453571.73276 &  510 \\ 
\hline 
\end{tabular}
\begin{tablenotes}[normal,flushleft]
\item $^a$ The uncertainty in the period, $\epsilon_P$, is given in units of $10^{-6}$ days.
\item $^b$ The best-fitting light curve template is given by $T$, where $T = 1$ -- 6 indicate
RRab stars with different shapes, and $T = 7$ indicates an RRc template.
\item Table \ref{tab_asas} is published in its entirety in the online edition; a portion is shown
here for guidance regarding its form and content.
\end{tablenotes}
\end{threeparttable}
\end{table*}

The intensity-mean magnitudes ranged from 10.2 to 13.8~mag, with a
median of 12.9~mag.  
In general, stars with fainter magnitudes showed more scatter in their light curves; the RMS values listed in Table~\ref{tab_asas} are good indicators of the degree of this scatter. Two of the stars, Z~For and 075127-4136.3, may be RRL first overtone pulsators (RRc) based on their  short periods and light curve behaviours. This supports the previous work of \citet{molnar16} who suggested the latter star is an RRc. We show below that the strengths of the hydrogen lines in our spectra further suggest that both stars are RRc.

\subsection{Equivalent Widths}    \label{ssec_ews}

For each spectrum, we measured the pseudo-equivalent widths (EWs) of the Ca~II~K,
H$\delta$, H$\gamma$, and H$\beta$ lines using the programs and
procedures described in \citet{layden94}.\footnote{Copies of the
Fortran code and supporting files for the EW-measurement and light curve fitting procedures are available at
\url{http://physics.bgsu.edu/~layden/ASTRO/DATA/EXPORT/progs.htm}. }
Briefly, the three hydrogen lines were each measured by fitting a
straight line across continuum band midpoints on the blue and red
sides of the feature as defined by the wavelengths shown in Table~5 of
L94 (after adjusting them for the star's radial velocity), and summing
the area enclosed between the stellar spectrum, the fitted continuum,
and the wavelength limits of the feature also shown in Table~5 of L94.
Because the H8 and H$\epsilon$ lines crowd the continuum around the
Ca~II~K line, we allowed continuum bands of a fixed width to shift and
seek a local maximum between broad limits (defined in Table~5 of L94),
and drew the continuum line between those mean points, as described in
Sec.~3.4.1 of L94.  Also, we measured the Ca~II~K line EW using a
narrower ($\lambda = 14$~\AA) feature band suitable for weak K lines,
and a wider ($\lambda = 20$~\AA) feature band suitable for strong K
lines.  In all, we measured equivalent widths on 600 useable stellar
spectra, 525 of which were of program RRL stars and 75 of which were
of the six EW standard stars discussed in the previous section 
(see Table~6 of L94).

These spectra were gathered on various nights over 18~months, enabling
us to test the long-term stability of our EW measurements. 
For a given EW standard star, the variance in EW measurements for a given spectral line was well
below 0.2 \AA\ in most cases, and we found no evidence of EW changes
over time.  We conclude that our observational system yields stable
and reliable EW measurements over the duration of our observing
program.

Because we will use the relationship between the EWs of Ca~II~K and
the hydrogen lines developed as Eqn.~7 of L94 to determine [Fe/H], it
is imperative that our present EWs have no systematic differences with
respect to those of L94.  Such differences could arise from the
different resolutions and detector quantum efficiency curves of the
respective spectrographs.  
For each spectral feature (Ca~II~K, H$\delta$, H$\gamma$, and H$\beta$), 
we plotted our observed EW against that star's standard EW from 
L94, and performed a linear least-squares fit
to obtain a slope and a zero-point coefficient.
For the Ca~II~K fit, we used EWs from the narrow
feature band for points with $W(K_{wide}) < 4.0$~\AA, and the wide
feature band for points with $W(K_{wide}) \geq 4.0$~\AA~ (following
L94).
In general, the linear fits did a good job of describing the observed
relationship across the range of EWs expressed by most of the RRL, with 
rms values of 0.26~\AA\ (Ca~II~K), 0.12~\AA\ (H$\beta$), 0.13~\AA\ (H$\gamma$), and 0.19~\AA\ (H$\delta$).
Inverting these linear relationships  
from the EW standard stars enabled us to convert our observed EWs of the RRL stars onto the standard
system defined by L94, and thereby utilize his Eqn.~7.

During each visit to a particular RRL star, we usually obtained three
spectra, and hence three sets of standardized EWs.  After deleting a
few EWs due to very low signal-to-noise spectra or obvious
contamination by cosmic rays, we averaged the remaining standardized EWs to obtain $W(K_o),
W(H\delta), W(H\gamma)$, and $W(H\beta)$, respectively, and computed
their standard deviations of the mean ($\epsilon$).  Table~\ref{tab_ews}
provides these values for our program RRL stars, along with an
estimate of the mean signal-to-noise ($S/N$)  
derived from the spectra in the continuum bands around the H$\gamma$ line, and the number of
spectra combined ($N_{sp}$) for each visit.  The table also contains the star name, the mean
heliocentric Julian Date of the visit, and its photometric phase, $\phi$,
predicted using the period in Table~\ref{tab_asas}.  We estimated the uncertainty in this phase,
$\epsilon_\phi$,  by recomputing the phase using our best period plus-and-minus
the period uncertainty from Table~\ref{tab_asas}. 
The table contains data from 525 useable spectra, combined
into mean values at 171 independent visits, for 89 individual RRL.

\begin{table*}

\caption{RRL Standardized Equivalent Widths}
\begin{threeparttable}
\label{tab_ews}
\begin{tabular}{cccccccccccccc}
\hline
ASAS~ID & HJD & $\phi$   & $\epsilon_\phi$  & $W(K_o)$    & $\epsilon_K$ &
$W(\delta)$  &  $\epsilon_\delta$ &                     $W(\gamma)$ & $\epsilon_\gamma$ &
$W(\beta)$ & $\epsilon_\beta$ &       $S/N$    & $N_{sp}$    \\

\hline
000602-3654.3 & 2455113.67921 & 0.112 & 0.005 &  1.68 & 0.08 &  6.26 & 0.12 &  5.93 & 0.13  & 5.71 & 0.10 &  34.8  & 3  \\
000602-3654.3 & 2455122.61268 & 0.491 & 0.005 &  2.78 & 0.12 &  4.25 & 0.08 &  3.52 & 0.13  & 4.07 & 0.17 &  24.7  & 4  \\
000602-3654.3 & 2455425.76933 & 0.739 & 0.006 &  3.05 & 0.05 &  4.01 & 0.22 &  3.46 & 0.09  & 3.96 & 0.11 &  28.0  & 3  \\ 
000602-3654.3 & 2455428.87690 & 0.698 & 0.006 &  2.66 & 0.10 &  4.31 & 0.23 &  3.66 & 0.04  &3.83 & 0.08 &   29.3  & 3  \\   
000621-3517.2 & 2455113.68797 & 0.930 & 0.128 &  5.51 & 0.03 &  3.97 & 0.17 &  3.99 & 0.11  & 4.40 & 0.16 &  16.3  & 3  \\     
000621-3517.2 & 2455122.62116 & 0.394 & 0.128 &  4.34 & 0.08 &  5.53 & 0.16 &  4.98 & 0.15  & 5.14 & 0.06 &  28.7  & 3  \\       
001543-5853.1 & 2455425.79257 & 0.892 & 0.009 &  2.37 & 0.01 &  4.30 & 0.01 &  4.41 & 0.01  & 4.02 & 0.01 &  30.7  & 1  \\         
001543-5853.1 & 2455431.83502 & 0.551 & 0.009 &  2.51 & 0.28 &  3.82 & 0.18 &  3.66 & 0.13  & 3.95 & 0.16 &  23.3  & 3  \\           
002418-7616.9 & 2455488.54552 & 0.532 & 0.202 &  2.73 & 0.11 &  4.65 & 0.27 &  3.90 & 0.03  & 4.08 & 0.23 &  24.1  & 3  \\             
... & ... & ... & ... & ... & ... & ... & ... & ... &  ... & ...  & ... & ... & ... \\
231648-4539.0 & 2455425.75349 & 0.508 & 0.042 &  3.40 & 0.13 &  4.15 & 0.14 & 3.48 & 0.12 &  3.63 & 0.05 &  33.3 &  3  \\ 
231648-4539.0 & 2455438.74129 & 0.022 & 0.042 &  2.54 & 0.19 &  6.11 & 0.04 & 5.36 & 0.06 &  5.11 & 0.04 &  29.2 &  2  \\ 
232246-4641.5 & 2455065.80512 & 0.693 & 0.010 &  4.73 & 1.50 &  3.56 & 0.69 & 3.69 & 0.22 &  4.02 & 0.20 &  11.5 &  3  \\ 
232246-4641.5 & 2455073.80513 & 0.143 & 0.010 &  3.12 & 0.27 &  6.64 & 0.12 & 6.13 & 0.19 &  6.12 & 0.11 &  25.6 &  4   \\
\hline 
\end{tabular}
\begin{tablenotes}[normal,flushleft]
\item Table \ref{tab_ews} is published in its entirety in the online edition; a portion is shown here for guidance regarding its form and content.
\end{tablenotes}
\end{threeparttable}
\end{table*}

\begin{figure}
\includegraphics[width=\columnwidth]{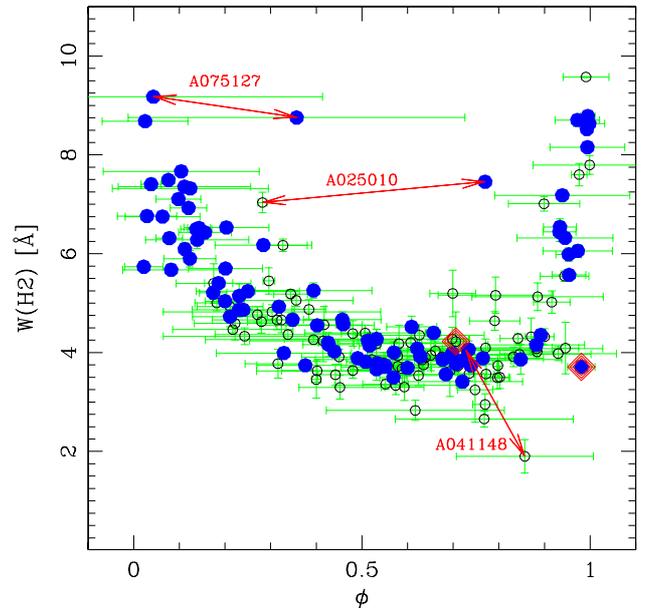}
\caption{The Balmer equivalent widths of our RRL spectra are plotted against
their photometric phases.  Solid and open circles indicate spectra
with $S/N \geq 20$ and $< 20$, respectively.  
Discrepant points include two possible c-type RRL (Z~For = 025010-2615.9 and 075127-4136.3) 
and an anomalously cool star (041148-3042.6).  Two spectra showing
evidence of Balmer emission are marked by concentric diamonds.}
\label{fig_h2phi}
\end{figure}

These equivalent widths provide us with a means for testing the phases
we derived in Sec.~\ref{ssec_asas} using the ASAS photometry.  
RRL are brightest when they are
hottest, and the EW of the Balmer lines is a good surface temperature
indicator \citep{smith95}.  In Figure~\ref{fig_h2phi} we plot $W(H2)$, the
average of the H$\delta$ and H$\gamma$ EWs, as a function
of photometric phase, $\phi$.  Most of the stars follow a pattern
similar to an RRL light curve, showing high surface temperature near $\phi = 0$ and cooling through $\phi \approx 0.8$.
The modest amount of scatter in this diagram indicates the stars' photometric phases are generally
correct.  

A few discrepant points are worth noting.  The stars Z~For and
075127-4136.3 have anomalously high $W(H2)$ values,  consistent with their being RRc as described previously. 
The spectra of 041148-3042.6 at one visit have extremely small Balmer
equivalent widths and extremely strong Ca~II~K lines, indicating an
anomalously cool star, while the spectra taken at the second visit
appear typical of RRL. Consequently, this star may not be an RRL.  During the process of measuring EWs, spectra
at two visits were noticed to have visual evidence of Balmer emission cores,
a well-known phenomenon in type-a RRL  \citep[e.g.,][]{preston64}.  This leads to
low EW measurements for the Balmer lines and an underestimate of the
star's [Fe/H].  The emission is obvious for RW~Hyi  
(the point at $\phi = 0.98$ in Figure~\ref{fig_h2phi}), but is less certain for 
UZ~Eri
(the point at $\phi = 0.71$).  With most of these outlier points
explained, we have enhanced confidence in the phases of of the
remaining points in Figure~\ref{fig_h2phi}.

Because our spectra were obtained in queue observing mode, it was not
practical to time the data acquisition to observe the star at the
optimal phase of $0.4 < \phi < 0.8$ (\citet{preston64}, L94).  Instead, our
spectra were obtained at random phases, and about 20\% were obtained
on the ``rising branch,'' $0.8 < \phi < 1.0$ (see
Figure~\ref{fig_h2phi}).  Rather than reject those spectra out of hand,
we aim to test whether spectra compromised by Balmer emission can be
identified and rejected, while the unaffected spectra can be retained.
Since most of our stars were visited twice at different phases, we can
calculate a preliminary [Fe/H] value from each visit using the mean
$W(H2)$ Balmer line width and the observed $W(K_o)$ line strength, along
with the coefficients from the first line of Table~8 in L94.  For each
pair, we denote the value from the spectrum with phase nearest to
$\phi = 0.4$ as [Fe/H]$_n$ and the one with the farther phase as
[Fe/H]$_f$.  We plotted the difference, [Fe/H]$_n - $ [Fe/H]$_f$,
against the phase farther from 0.4; stars with far-phases between 0.0
and 0.8 have a fairly flat distribution centered on zero with a
standard deviation of 0.20 dex, while stars with far-phases between
0.8 and 1.0 (the rising branch) scatter more widely.  Several of these
outliers include the two visits with Balmer emission mentioned above,
and the anomalously cool star 041148-3042.6.  We settled on the following
prescription for handling the [Fe/H] values from spectra taken on the
rising branch: for stars with paired [Fe/H] values deviating by 0.4
dex ($2\sigma$) or more from the mean, we did not use spectra taken
with phases between 0.8 and 1.0 (36\% of the rising branch spectra),
while spectra deviating by less than 0.4 dex received half-weight
(64\%).

With the visits flagged to have full, half, or no weight, we refined
the [Fe/H] values by applying the interstellar Ca~II~K and H$\beta$
metal-line contamination corrections described in L94.  For the
latter, we calculated the metal line strength, $W(Z_{\beta})$, using
Eqn.~3 of L94 and computed the corrected H$\beta$ equivalent width as
$W(H\beta_z) = W(H\beta) - W(Z_{\beta})$ (see column~4 of Table~\ref{tab_findat}). For the former, we used the Galactic model of
$W(K_{int})$ described by \citet{beers90} (see column~3 of Table~\ref{tab_findat}) and estimated an uncertainty in this value using the
relation
\begin{equation}
\epsilon_{Kint} = 0.018 ~\rm{ln}~ |Z| + 0.060
\end{equation}
derived from the 1-$\sigma$ error values in Table~II of
\citet{beers90}, where $|Z|$ is the distance from the Galactic plane
in kpc.  To obtain a preliminary heliocentric distance to each star,
and hence $|Z|$, we assumed the absolute magnitude relation of
\citet{cacciari13} along with the
intensity-mean apparent magnitude listed in Table~\ref{tab_asas} and
the interstellar reddening listed from Table~\ref{tab_coords}.  We used
the corrected calcium equivalent width, $W(K) = W(K_o) - W(K_{int})$,
and the combined Balmer line strength, $W(H3) = [W(H\delta) + W(H\gamma)
+ W(H\beta_z)]/3$, along with the coefficients in line~3 of Table~8 in
L94 to get the final [Fe/H] estimate for each visit.

We estimated the uncertainty in the [Fe/H] for each visit by
recomputing [Fe/H] using three variations, and adding in quadrature
their differences from the best [Fe/H] value.  The first variation
changed the assumed $W(K)$ by adding the observational error
$\epsilon_K$ from Table~\ref{tab_ews}; the second variation changed
the assumed $W(K)$ by adding the error in the interstellar correction
$\epsilon_{Kint}$; and the third variation changed the assumed $W(H3)$ by
the combined errors $\epsilon_\delta$, $\epsilon_\gamma$, and
$\epsilon_\beta$ from Table~\ref{tab_ews}.

For a given star, we combined the [Fe/H] values from each visit using
a weighted mean, in which the weights were the inverse-squares of the
uncertainties described above, multiplied by 0.5 or 0.0 for rising-branch phases.  
These values are listed in column~5 of Table~\ref{tab_findat}, along with the weighted error ($\epsilon_1$), the
standard deviation of the mean of the [Fe/H] values from each visit around the mean
($\epsilon_2$), and the effective number of visits used ($N_{vis}$,
where the half- and zero-weighted rising-branch visits contribute 0.5
and 0 to the count).  The median value of $\epsilon_2$ is 0.06 dex;
this should describe the internal error in a typical star's [Fe/H],
that is, its precision in the metallicity ranking within our study.  A
larger systematic uncertainty may shift our metallicity ranking with
respect to those of other studies.

\begin{table*}
\caption{Final Spectroscopic Data}
\begin{threeparttable}
\label{tab_findat}
\begin{tabular}{ccccccccrrrc}
\hline
ASAS~ID & Alt.~ID &  $W(K_{int})$ &  $W(H\beta_z)$ & [Fe/H] & $\epsilon_{1}$  & $\epsilon_{2}$  &  $N_{vis}$ 
  & $v_{CM}$  & $\epsilon_{v_{CM}}$   & $N_{fit}$  & Comment \\
\hline
000602-3654.3 &   WW~Scl  & 0.19 & 0.11 & --2.23 & 0.03 & 0.03 & 4.0 &     31 &  18 & 13 & ... \\
000621-3517.2 & CD~Scl  & 0.19 & 0.30 & --1.35 & 0.05 & 0.06 & 1.5 &   --83 &  34 &  6 & ... \\  
001543-5853.1 &   UZ~Tuc  & 0.21 & 0.10 & --2.33 & 0.04 & 0.01 & 1.5 &   77 &  19 &  6 & ... \\   
002418-7616.9 &    T~Hyi & 0.27 & 0.11 & --2.23 & 0.07 & 9.99 & 1.0 &     --37 &  27 &  3 & ... \\      
002443-6949.7 &     ...  & 0.15 & 0.46 & --1.13 & 0.05 & 0.22 & 2.0 &   139 &   16 &  6 & ... \\       
002843-4400.4 & NSV00174  & 0.19 & 0.26 & --1.46 & 0.15 & 0.22 & 1.0 &  --165 &  17 &  6 & ... \\     
  ...   &  ... & ... & ... & ... & ...  & ...  &   ...  & ...  & ...  & ...  & ...  \\
221843-5652.4 &      ...  & 0.24 & 0.12 & --1.82 & 0.10 & 0.03 & 1.0   & --212 &  58 &  5 & ... \\
231648-4539.0 &      ...  & 0.20 & 0.14 & --2.05 & 0.05 & 0.01 & 2.0   &   61 &  21 &  5 & ... \\
232246-4641.5 & NSV14530  & 0.19 & 0.17 & --1.58 & 0.13 & 0.02 & 2.0  &   --18 &  19 &  7 & ... \\
\hline 
\end{tabular}
\begin{tablenotes}[normal,flushleft]
\item Table \ref{tab_findat} is published in its entirety in the online edition; a portion is shown here for guidance regarding its form and content.
\end{tablenotes}
\end{threeparttable}
\end{table*}

Four stars of our sample (RX Col, VW Dor, UU Hor, SV Vol) are in common with L94.  The mean difference in [Fe/H] in the sense L94-ours is 0.06 with $\sigma = 0.05$.  This small difference suggests that we have successfully placed our measurements on the L94 scale to within the errors in our measurements.

Only two stars in our sample have been previously measured by high-resolution spectroscopy, XX~Pup and VW~Scl, for which we obtained [Fe/H]$=-1.59$ and $-1.55$, respectively.   \citet{fernley97} obtained $-1.33$ for XX~Pup, and \citet{pancino15}, obtained $-1.22\pm0.10$ for VW~Scl.  At face value, this suggests that our measurements of [Fe/H] are too low by about 0.3 dex.  This amount of disagreement is not surprising, however, in light of the differences between different high-resolution studies.  For example, the 4 stars in common between \citet{fernley97} and \citet{pancino15}, have a mean deviation of 0.23 dex with $\sigma = 0.42$.  The differences between high-resolution measurements for RRL have been discussed by \citet{chadid17}, who pointed out that many high-resolution [Fe/H] scales transform well to the L94 scale.  Due to the lack of consensus on the [Fe/H] scale, we use L94 scale below in our discussion of the metallicities and kinematics of local RRLS.         

\subsection{Radial Velocity Measurements}    \label{ssec_vrad}

We used the Fourier cross-correlation method as implemented in the
IRAF task FXCOR to measure the wavelength shift between an object
spectrum of unknown radial velocity (RV) and a template spectrum of
known velocity. At the low-resolution of our spectrograms and in the window 3750--5350~\AA, the cross-correlation method is most sensitive to the positions of the Balmer lines, whose strengths primarily depend on the effective temperatures of stars at the time of observation.  For templates, we used the synthethic spectra calculated by \citet{munari05} for temperatures from 5500 to 8500 K in steps of 500K, a surface gravity of log g$= 2.5$, and [Fe/H]$= -1.0$.  We also included a metal-poor and a metal-rich template ([Fe/H]$ = -2.5$ and $ +0.0$) at 6500 K, and log g$= 2.5$, because of the wide [Fe/H] range of our sample ($\sim -2.6$ to $ \sim +0.3$).  This yielded nine templates at a resolution of 1~\AA.  Next, we correlated the RRL spectra against each of the nine synthetic templates, and took the weighted mean RV and its error as the RV and uncertainty for each spectrum.  The velocity stemming from the best match between the stellar spectrum and the template was therefore given the most weight.  To check whether it was essential to match the template with the stellar [Fe/H], we also computed the differences in velocity given by each of the three templates of different [Fe/H]. This revealed that there were small offsets in the RV's given by these templates, which averaged $< 10$ km~s$^{-1}$, but with no correlation with stellar [Fe/H].

The radial velocity curves of RRL depend on the mode of pulsation and the amplitude of of the light variation.  \citet{sesar12} combined the RVs from high resolution spectra of several bright RRab to obtain template radial velocity curves at several wavelengths.  We chose the curve based on the H$\gamma$ line \citep[see Figure~1 of][]{sesar12} as it was near the center of our spectrum, and we scaled its RV amplitude using the
 photometric amplitude ($\Delta V = V_{min} - V_{max}$) of each RRL
from Table \ref{tab_asas} and Equation~5 of \citet{sesar12}.
For each RRL, we performed a least-squares fit between the observed RV
and phase data, and the scaled RV template, yielding a systemic, center-of-mass velocity for the pulsating star, $v_{CM}$.  \citet{sesar12} noted an uncertainty of $\sigma_{v_\gamma} \approx 13$ km~s$^{-1}$  in $v_{CM}$ due to the uncertainty in the phase on the template at which the systemic velocity occurs. 
RRc variables have much smaller RV amplitudes than RRab, and the precise form of RV curve is less critical.  For the two RRc in our sample, we used the same RV curve as \citet{duffau06}, which is based on curves of T Sex and DH Peg. 

\begin{figure}
\includegraphics[width=\columnwidth]{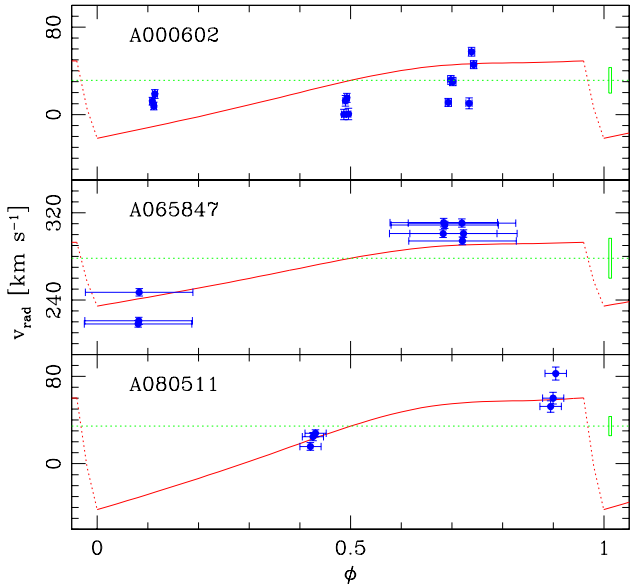}
\caption{Radial velocity fits for three RRL: 000602-3654.3 (top), 065847-4926.0 (middle), and 080511-2120.4 (bottom).  In each panel, the blue
symbols mark the observed radial velocities at the phases predicted by ASAS photometry.  The red curve
is the best-fit radial velocity template, scaled from each star's
photometric amplitude.  The green dotted line is the center-of-mass
velocity of the best-fit RV curve, and the box near $\phi = 1$
indicates the total estimated uncertainty in $v_{CM}$.}
\label{fig:rvcurves}
\end{figure}

Figure~\ref{fig:rvcurves} shows examples of fits for three selected stars.  The top panel shows a case where there were a large number of individual spectra in four separate visits, but poor agreement between the RVs from the different visits.  The middle panel shows a case with better agreement between the three visits, but relatively large uncertainties in the photometric phase.  The bottom panel shows an excellent fit.  The amount of scatter  of the individual RV measurements around the best-fit template (specifically, the standard deviation of the mean) is a second component of the overall uncertainty in our estimate of $v_{CM}$.

A third component of this uncertainty comes from the uncertainty in our photometric phases, $\epsilon_\phi$.  We assessed this contribution for each RRL by 
refiting the template to the radial velocity data after shifting it by $+\epsilon_{\phi}$ and again by
$-\epsilon_{\phi}$; the resulting velocity shifts were usually between
3 and 10 km s$^{-1}$, though larger cases occurred as shown in the middle panel of Figure~\ref{fig:rvcurves}.  

To determine the overall uncertainty in a star's center-of-mass radial velocity, $\epsilon_{CM}$, we added in quadrature 
this velocity shift, the standard deviation of the mean of the
individual velocity-phase observations around the best-fit RV curve, and $\sigma_{v_\gamma}$ from the template.  The  final
values of $v_{CM}$ and its error are shown in Table~\ref{tab_findat},
along with the number of spectra used in the fit, $N_{fit}$.  Errors
in $v_{CM}$ ranged  from 14 to 58~km~s$^{-1}$ with a median of 21~km~s$^{-1}$.  Given the small number of independent visits involved in
fitting each star and the statistical ``jitter'' this would introduce,
it is reasonable to believe that the true uncertainty in the RV of
most stars in our sample is of order 20--25 km~s$^{-1}$.

A comparison of our radial velocities with ones in the literature gives mixed results.  For the 4 stars in common with L94, the mean difference in the sense L94-ours is 8~km~s$^{-1}$ with $\sigma = 10$~km~s$^{-1}$, and each diffence is smaller than the errors combined in quadrature.  The velocities of VW Scl and XX Pup were measured by \citet{britavskiy18} and \citet{fernley97}, respectively, from high-dispersion spectrograms.  The mean difference between their velocities and ours is 14~km~s$^{-1}$, but with a huge scatter ($\sigma= 64$~km~s$^{-1}$).  Another check on our velocities is provided by the recent Data Release 2 measurements by the \textit{Gaia} satellite.  Since the \textit{Gaia} velocities are not corrected for the phase of observation, we considered only the mean radial velocities measured from 5 or more \textit{Gaia} observations.  For the 25 stars in common, the mean difference in the sense \textit{Gaia}-our velocities is 15.8~km~s$^{-1}$ with $\sigma=21.9$.  This is reasonably good agreement since the mean of adding the errors in quadrature is 18.5~km~s$^{-1}$.  Because one of the above comparisons produced poor results, we caution that the errors of our velocity measurements may be under estimated.                 

\begin{figure}
\includegraphics[angle=270,width=\columnwidth]{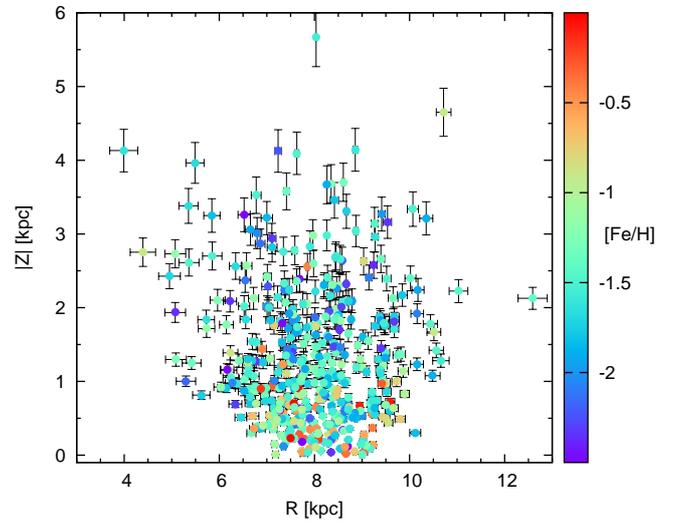}
\caption{The RRL are plotted with cylindrical coordinates. For many stars, the error bars are smaller than the plotted points.}
\label{fig:RZ}
\end{figure}

\begin{figure}
\includegraphics[angle=270,width=\columnwidth]{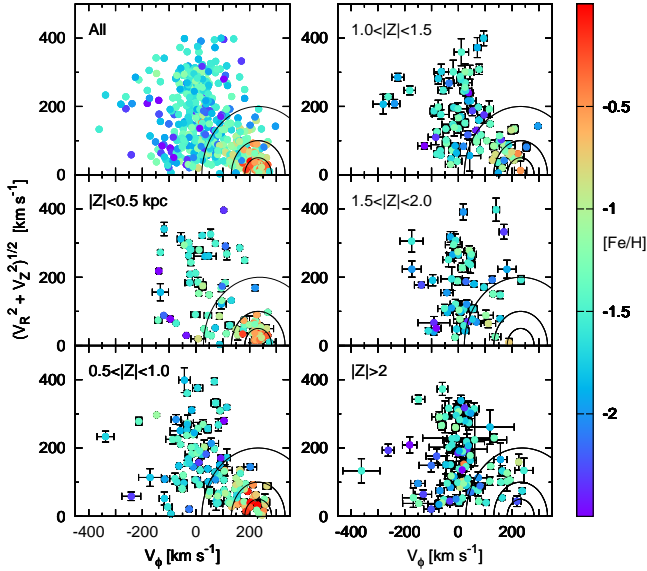}
\caption{The RRL in zones $|Z|$ are plotted in Toomre diagrams for velcities in the galactic rest frame, $V_{LSR}= 232$\, km\, s$^{-1}$.  The stars are color coded by $[Fe/H]$.  For reference, semi-circles are drawn around $V_{LSR}$ at $50$, $100$, and 210 \, km\, s$^{-1}$, the halo-disc dividing line of \citet{helmi18}.  To preserve clarity, error bars are not plotted in the upper left diagram. For several stars in these diagrams, the error bars are smaller than the points.}   
\label{fig:toomre}
\end{figure}

\section{Kinematics of Local RR Lyrae Stars}
\label{sec:kinematics} 
\subsection{The Sample}

To investigate the kinematics of the local RRL and their dependence on [Fe/H], we have compiled a catalogue of radial velocity and [Fe/H] measurements for 464 stars.  We adopted the L94 [Fe/H] scale because of the large number of stars measured by L94 and because \citet{chadid17} have shown that while several [Fe/H] scales based on high-resolution spectroscopy disagree among themselves, each transforms well to the L94 scale.  The L94 scale is anchored in the older \citet{zinn84} scale for globular clusters for [Fe/H]$_{ZW}\leq -0.99$, but at higher metallicities it is based entirely on measurements of field RRLS transformed to the ZW scale.  We caution, therefore, that a transformation between the ZW scale and another globular cluster scale may not be applicable to the whole range of the L94 scale.

Our primary source for the catalogue was the earlier one by \citet{dambis13}, who had compiled measurements of [Fe/H] on the L94 scale and radial velocities from numerous authors. We deleted from their catalogue the few stars more distant than 6 kpc and also BB Vir and BI Tel because the first is a blend and the second is an Anomalous Cepheid \citep{muraveva18}.  We also deleted two stars measured here, 041148-3042.6 and 083412-6836.1, because they may not be RRL (see above). 

The stars in our catalogue are listed in Table~\ref{tab_inputcat}, where possible the name in the variable star convention is listed from the Simbad database, along with the equatorial coordinates (J2000.0).  For the stars not measured here, a major source for the periods (P) and amplitudes in the V pass-band (Amp) of the variables was \citet{skarka14}.  Many of the designations as type ab or c, <V>, and the interstellar extinctions in the V pass-band (Vext) were taken from \citet{dambis13}.  Other sources for these data were the references listed by \citet{magurno18}, \citet{muraveva18}, and the Simbad database.  

The column labelled `m-src' in Table~\ref{tab_inputcat} describes the source of the metallicity measurement.  A  `1' in this column indicates the [Fe/H] value was from L94, the present paper, or the mean of both.  These measurements are from low-resolution spectrograms following the L94 technique.  A `2' as the first digit indicates the measurement was taken from \citet{dambis13}, and it is based on a transformation to the L94 scale.  A `3' indicates that we have tranformed one or more measurement from high-resolution spectroscopy to the L94 scale.  These linear transformations used the stars in common between L94 and each high-resolution study and the [Fe/H] values listed by \citet{magurno18}, which are the high-resolution results re-scaled to the same solar abundance.  The transformations contained from 8 to 26 stars, and their rms values varied from 0.10 to 0.16 dex.  Where possible, the values transformed to the L94 scale were averaged with each other and with the L94 values.  A good example of the utility of this procedure is provided by the very metal-poor star X Ari, for which the results from high-resolution spectroscopy disagree \citep[see also][]{chadid17}.  The 6 measurements compiled by \citet{magurno18} range from -2.19 to -2.74, and have a mean, <[Fe/H]>, of -2.52 and $\sigma = 0.19$.  After transforming the individual measurements to the L94 scale, the range narrows to -2.34 to -2.58 with <[Fe/H]>$= -2.44$ and $\sigma = 0.11$.  The L94 value is -2.40, and the value in Table~\ref{tab_inputcat} is the average of this and the 6 transformed values. The column labelled `N'  in the table is the number of independent studies that contributed to the listed mean.  Thirty-five stars have 3 or more measurements, and average of their $\sigma$'s is 0.07.  This is essentially the same as the error obtained with the L94 technique (see above), and it is our estimate of the error in the [Fe/H] rank from one measurement.    

A few stars, mostly type c, with m-src~$ = 4$ have values of [Fe/H] from high-resolution spectroscopy that could not be transformed to the L94 scale.  We have adopted the values of [Fe/H] listed in \citet{magurno18} for these stars.  Our experience with the other results from high-resolution spectroscopy suggests that these measurements are probably not more than 0.3 dex from the L94 scale.
 
The columns labelled $v_{CM}$ and $\epsilon_{CM}$ in Table~\ref{tab_inputcat} lists values of the adopted systemic velocity and its error.  The column labelled `v-src' describes the source of the measurements.  A `1' in this column indicates that the velocity and its error are from the literature or this paper.  Many of the literature values were taken directly from the \citet{dambis13} compilation, but first we corrected the data for a few stars that appear to have been miss copied from \citet{jeffery07}.  We also adopted the velecities that \citet{jeffery07} measured by fitting a radial velocity curve to one measurement, but without quoting an error for this procedure.  On the basis of the small errors that they quote for 2 or more observations, we have assigned  $\epsilon_{CM} = 5$~km~s$^{-1}$ to a single measurement.  We also adopted the velocities that \citet{britavskiy18} and \citet{sneden17} measured by fitting radial velocity curves to their measurements of high-resolution spectrograms.  Other literature values were taken from the Simbad database, but only if a velocity curve was fit to the measurements made at specific phases.   The velocities with v-src~$= 2$ are from the mean of 5 or more \textit{Gaia} observations without the fit of a velocity curve.  These values were adopted when a previous measurement was based on a few low-resolution spectrograms, which resulted in a relatively large $\epsilon_{CM}$.  

The above procedures produced a catalogue of 466 RRL with measured metallicities and radial velocities.  The parallaxes and proper motions of this sample were downloaded from the \textit{Gaia} DR2 Archive, and then examined for reliability and consistency.  The DR2 data for RR Lyra itself is obviously spurious, which is attributable to an incorrect measurement of its G magnitude \citep{muraveva18}.  For this one star we adopted the \textit{Gaia} DR1 results, which are based on the TYCHO-GAIA astrometric solution.  Two other stars in our initial catalogue, AT And and VV Peg, have large values of the `unit weight error' (UWE) and the `renormalized unit weight error' (RUWE) \citep[see][]{lindegren18}, which indicate that their DR2 data are untrustworthy. \footnote{In addition, see `Re-normalising the astrometric chi-square in Gaia DR2' which is available at www.cosmos.esa.int/web/gaia/public-dpac-documents.} These stars were removed from the sample. For the remaining 463 stars, the reliability criteria `visibility\_periods\_used', RUWE, and UWE are $\ge 6$, $\le 1.61$, and $\le 2.38$, respectively.  Only the RUWE criterion is above the recommended cutoff ($\le 1.4$) for good astrometry.  We relaxed this criterion slightly because RUWE depends on the magnitudes and colors measured by Gaia, which of course for RRL vary with phase.  None of the stars making up the 5\% of the sample that have RUWE $> 1.4$ have UWE values outside the range of good astrometry.
    
The DR2 proper motions for bright stars (\textit{G} < 13) are affected by a global rotation bias that varies with position and magnitude.  This was removed by computing the corrections for each star using the formulae and coefficients given in \textit{Gaia} DR2-known-issues.\footnote{See www.cosmos.esa.int/web/gaia/public-dpac-documents.}  We applied them in full to the stars with \textit{G}$\le 11.0$, with a factor \textit{f}($=(13.0-G)/2.0$) if $11.0 < G \le 13.0$, and with no correction ($f=0$) if $G > 13.0$.  These corrections are very small with respect to the proper motions and applying them change the space velocities of the stars by $< 1$~km~s$^{-1}$.

The errors in the distances of the stars obviously affect both the positions of the stars in the Galaxy and their tangential velocities.  The distances to our sample can be deduced from their parallaxes and from the absolute magnitude-metallicity relation for RRL.  For the latter, we adopted the relationship that \citet{muraveva18} obtained using the \textit{Gaia} DR2 parallaxes for 23 very nearby RRL:

\begin{equation}
M_{V}=0.26[Fe/H]+1.04.
\label{eq:Mv}
\end{equation}

It is well documented that the \textit{Gaia} parallaxes suffer from an offset in the sense that the measured parallaxes are too small. \citet{muraveva18} added an offset of 0.057 mas to the DR2 parallax when deriving equation~(\ref{eq:Mv}), and the same value is adopted here.  They also showed that equation~(\ref{eq:Mv}) yields a distance to the Large Magellanic Cloud that is in excellent agreement with independent measurements.     

\begin{table*}
\caption{The Sample of Local RR Lyrae Stars}
\begin{threeparttable}
\label{tab_inputcat}
\begin{tabular}{rrrrrrrrrrrrrr}
\hline
Name & RA (J2000) & Dec (J2000) & Period & Amp & Type & <V> & Vext & [Fe/H] & m-src & N & $v_{cm}$ & $\pm$ &  v-src\\
\hline
SW~And &  5.929507 &  29.400927 &  0.44226  &  0.94 &  ab  &  9.71 &   0.11 &  -0.38 &  3 &  7 &  -21.0  &   1.0 & 1\\
XX~And & 19.364548 &  38.950428 &  0.72277  &  0.99 &  ab  & 10.69 &   0.00 &  -2.01 &  1 &  1 &    0.0  &   1.0 & 1\\
XY~And & 21.676867 &  34.068539 &  0.39871  &  1.00 &  ab  & 13.68 &   0.14 &  -0.92 &  1 &  1 &  -64.0  &  53.0 & 1\\
ZZ~And & 12.395349 &  27.022041 &  0.55455  &  0.90 &  ab  & 13.08 &   0.12 &  -1.58 &  1 &  1 &  -13.0  &  53.0 & 1\\
BK~And & 353.775177 & 41.102901 &  0.42160  &  0.71 &  ab  & 12.97 &   0.34 &   0.08 & 1  & 1  & -17.0 &    7.0 & 1\\ 
\hline
\end{tabular}
\begin{tablenotes}[normal,flushleft]
\item m-src describes the source of the [Fe/H] value.  1 = L94 and/or this paper; 2 = \citet{dambis13}; 3 = includes values transformed to L94 scale from high-resolution spectroscopy; 4 = not transformed high-resolution value.
\item v-src = 1 velocity from literature, = 2 from Gaia DR2.
\item Table \ref{tab_inputcat} is published in its entirety in the electronic edition of the Journal; a portion is shown here for guidance regarding its form and content.
\end{tablenotes}
\end{threeparttable}
\end{table*}

The reciprocal of a measured parallax will yield an accurate distance as long as the ratio of the parallax to its error ($\tilde{\omega}/\sigma_{\tilde{\omega}} )\gg 1$.  The distances obtained from equation~(\ref{eq:Mv}) and the reciprocal of the \textit{Gaia} parallax with the added offset are in good agreement for nearly all stars with $\tilde{\omega}/\sigma_{\tilde{\omega}} \ge 15$ ($\sim 7\%$ distance error), but become less so at smaller values, and poor at $\le 6$ in the sense that the distances from the parallaxes are too large.  The distances obtained from the absolute magnitude [Fe/H] relation are expected to have precisions of $5-7\%$ in the absence of large reddening or photometric uncertainties \citep[e.g.,][]{vivas06a}. Consequently, we have adopted the distances given by equation~(\ref{eq:Mv}) for 251 stars with $\tilde{\omega}/\sigma_{\tilde{\omega}} < 15$ and assumed a relative error of 7\%.   For 214 other stars this ratio is $\ge 15$, and their distances and errors were calculated from the \textit{Gaia} parallax data.  Because TX Car has a very large and uncertain interstellar extinction, we used its parallax distance even though $\tilde{\omega}/\sigma_{\tilde{\omega}} = 11.74$.  This methodology avoids where possible the uncertainty due to interstellar extinction, and it also avoids the inflation of the tangential velocities of the stars that would stem from adopting distances that are too large from the parallaxes.
     
The distances of the stars from the Sun ($d_{\odot})$, their rectangular coordinates, X, Y, and Z, their velocities in cylindrical coordinates, $V_{R}$, $V_{\phi}$, and $V_{Z}$, their angular momenta about the Z axis ($L_Z$) and perpendicular to the Z axis ($L_{\perp} = (L_{X}^2 + L_{Y}^2)^{-1/2}$), and the sum of their kinetic and potential energies ($E_{tot}$) are listed in Table~\ref{tab_kine} \footnote{X, Y, and Z point, respectively, from the galactic center through the Sun, in the direction of galactic rotation, and toward the North Galactic Pole (NPG).  $V_{R}$ is the velocity in the plane from the Z axis, $V_{\phi}$ is the azimuthal velocity, positive in the direction of galactic rotation, and $V_{Z}$ is the velocity out of the plane, positive toward the NGP}  To calculate these quantities, we adopted 8.2 kpc for the Sun's distance from the galactic center, a solar motion ($U_{\odot},V_{\odot},W_{\odot}$) = (11.1,12.24,7.25) \,km \,s$^{-1} $ \citep{schonrich10}, and 232 \,km \,s$^{-1} $ for the velocity of the local standard of rest (LSR).   With these quantities, the LSR has positive $L_{Z} = 1902$ kpc~km~s$^{-1}$. To calculate the potential energy, we used model 2 from \citet{pouliasis17} for the gravitational potential, which yields $230$\,km \,s$^{-1}$ for the circular velocity at 8.2 kpc.  With this potential and the above velocity, the LSR has $E_{tot} = -1.72 \times 10^5\ \rm{km}^{2}\ \rm{s}^{-2}$.

\begin{table*}
\caption{The Galactic Positions and the Kinematics of the Stars}
\begin{threeparttable}
\label{tab_kine}
\begin{tabular}{crrrrrrrrrrrrrrrr}
\hline
Name  & $d_{\odot}$ & $\pm$  &  X  &  Y &  Z & $V_R$ & $\pm$ & $V_{\phi}$ & $\pm$ & $V_Z$ & $\pm$& $L_Z$ & $\pm$ & $L_{\perp} $&  $\pm $ &  $E_{tot} $\\
\hline
SW~And  & 0.54 & 0.04 & 8.4 & 0.4 & -0.3 & -40 &  3 & 218 &  1 & -20 &  4 & 1831 &  12 &  194 &  29 & -1.72 \\
XX~And   & 1.33 & 0.10 & 9.0 & 0.9 & -0.5 & 244 & 16 & -74 & 18 & -145 &  9 &  -671 & 163 & 1174 &  84 & -1.50 \\
XY~And   & 3.53 & 0.25 & 10.2 & 2.3 & -1.7 &  75 & 32 &   5 & 38 & -68 & 26 &   49 & 395 &  587 & 270 & -1.79 \\
ZZ~And  &  2.93 & 0.20 & 9.5 & 2.0 & -1.7 & 243 & 31 & -175 & 43 & -185 & 34 & -1690 & 418 & 1406 & 301 & -1.26 \\
BK~And  & 1.99 & 0.14 & 8.8 & 1.8 & -0.7 &  71  & 4 & 193 &  6 & -16 &  3 & 1727 &  58 &  163 &  21 & -1.72 \\
\hline
\end{tabular}
\begin{tablenotes}[normal,flushleft]
\item The units are kpc, km~s$^{-1}$, kpc~km~s$^{-1}$, and $10^5$ km$^{2}$~s$^{-2}$ for the coordinates, velocities, angular momenta, and total energies, respectively.
\item Table \ref{tab_kine} is published in its entirety in the electronic edition of the Journal; a portion is shown here for guidance regarding its form and content.
\end{tablenotes}
\end{threeparttable}
\end{table*}

In Fig.~\ref{fig:RZ}, the cylindrical coordinates of the star are plotted with different colors representing their metallicities. About $99\%$ of the stars lie within the cylinder, roughly centered on the Sun, with $|Z| \le 4$ kpc and $4 \le R \le 12$ kpc ($R \equiv (X^2+Y^2)^{1/2}$).  The sample is undoubtedly biased towards brighter apparent magnitudes, and this and the large extinction near the galactic plane explains the fan shaped distribution of the stars in Fig.~\ref{fig:RZ}.  The metal-rich RRL ($[Fe/H] > -1$) are largely confined to $|Z|< 1$ kpc; consequently, the bias against stars near plane impacts their number most strongly.   

\subsection{The RRL Discs}

Beginning with the pioneering study of \citet{preston59}, numerous investigators have shown that the metal-rich RRL have the kinematic properties of a disc population. While the majority of these disc RRL resemble kinematically the thick disc population \citep[e.g.,][]{dambis13}, the evidence that some of belong to the thin-disc has mounted as the data have improved \citep{layden95a, maintz05,liu13,marsakov18b,marsakov18a}.  Since the \textit{Gaia} satellite has provided new and improved astrometric data, we examine this question again using our sample, which is somewhat larger than previous ones.

In the top left diagram of Fig.~\ref{fig:toomre}, we have plotted the entire RRL sample in the Toomre diagram.  One sees immediately that the majority of the stars with [Fe/H]~$< -1$ scatter widely over diagram, and much more so than the more metal-rich stars, which cluster within $\sim 100$\,km \,s$^{-1}$ of $V_{LSR}$.  These are the long-recognized halo and disc populations of metal-poor and metal-rich RRL, respectively. We have also plotted in Fig.~\ref{fig:toomre}, Toomre diagrams for 5 zones in |Z|, which illustrate the changes in kinematics and in [Fe/H] with distance from the plane. Thin- and thick-disc stars are commonly defined by their pecular velocities with respect to the LSR ($V_{pec}=(V_{R}^2 + (V_{\phi} - 232)^2 + V_{Z}^2 )^{1/2}$, and here we adopt $V_{pec} \le 50$ and $70 < V_{pec} < 180$\,km \,s$^{-1} $ for the thin- and thick-discs, respectively, following the discussion in \citet{bensby14}.  The ratio of the so-defined numbers of thin to thick stars vary as 21/18, 11/33, 2/24, 1/7, and 1/13 in the 5 zones of increasing |Z| portrayed in Fig.~\ref{fig:toomre}.

The observational errors are not responsible for placing some RRL in the thin-disc region of Fig.~\ref{fig:toomre}, for the majority of the 35 thin-disc stars have $|50-V_{pec}|/\sigma_{V_{pec}} > 2$.  The distances of $77\%$ of these stars were measured from well-determined parallaxes, and consequently, their space velocities are independent of the instellar extinction.  Because the samples of disc stars suffer from incompleteness, particularly near the galactic plane (see Fig.~\ref{fig:RZ}), and from contamination from the halo population, which on account of being both old and metal-poor has a higher frequency of RRL compared to the disc populations, it is not possible to measure reliably the scale heights of the discs.  The falloff with |Z| of the number of stars in the thin-disc region is in qualitative agreement with the scale height of $\sim 300$ pc for the old thin-disc \citep{bland-hawthorn16}.

The thin-disc and thick-disc samples of RRL overlap in [Fe/H] (see Fig.~\ref{fig:toomre}), but on average the thin-disc sample is more metal-rich.  For example, in the |Z| $\le 0.5$ region, where both discs are abundant in our sample, the kinematcally defined thin- and thick-disc samples have mean [Fe/H] $= -0.68$ and $-1.34$, respectively.  In the whole sample of 36 RRL in the thin-disc region, only one (T Hyi) is very metal-poor ([Fe/H] $<-2$), and because of its relatively large |Z|, 2.3 kpc, it is likely an interloper from another population. But 7 of the 95 thick-disc stars in our sample have $-2.33 \le $ [Fe/H] $\le -2.00$.  These stars and T Hyi may be related to the very metal-poor disc stars recently discussed by \citet{sestito19}.

\begin{figure}
\includegraphics[angle=270,width=\columnwidth]{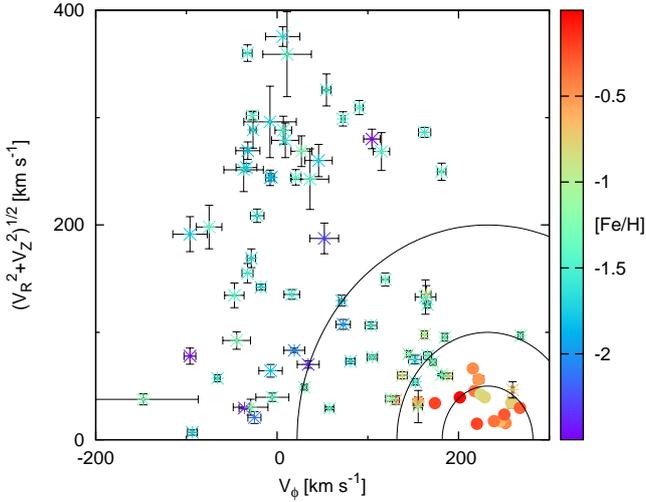}
\caption{The low (solid circles) and high (asterisks) [$\alpha$/Fe] groups in the Toomre diagram.  The semi-circles are drawn at $V_{pec}= 50, 100,$ and $210$\,km \,s$^{-1}$. For some stars, the error bars are smaller than the points.} 
\label{fig:alpha}
\end{figure}

Another discriminant between thin- and thick-disc stars near the Sun is the ratio of the abundances of the $\alpha$ elements to Fe, which at the same [Fe/H] is significantly higher than solar in thick disk stars and near solar in thin-disc ones \citep[e.g.,][]{haywood13,bensby14}.  For 96 field RRL, \citet{magurno18} have recently compiled and re-scaled to the same solar abundance the measurements in the literature of [Mg/Fe], [Ca/Fe], and [Ti/Fe] from high-resolution spectroscopy.  Fig. 10 in \citet{magurno18}, shows the dependence of these ratios on [Fe/H] and also illustrates the scatter in the measurements.  Eighty-six stars of their sample are in our catalog, and for these stars we averaged where possible the abundance ratios that \citet{magurno18} list from multiple sources.  On the basis of plots similar to their fig. 10, but with the mean abundance ratios, we divided our sample into low and high $\alpha$/Fe groups by making cuts at [Mg/Fe],[Ca/Fe] and [Ti/Fe] = 0.25, 0.1, and 0.1, respectively. Because not all 3 abundance ratios were measured for every star, we adopted the following procedure for assigning stars to a group.  For a star to be in the low $\alpha$ group, no more one of the above ratios could be above its cut if all 3 ratios were measured, and none above the cut, if $<3$.  Fig.~\ref{fig:alpha} shows the Toomre diagram for the so-defined low and high $\alpha$ groups.  Eleven of the 15 stars in the low $\alpha$ group and 0 of the 71 stars in the high $\alpha$ group have $V_{pec} \le 50$.  Hence, many of the RRL that are kinematically thin disc stars also have the thin-disc charactistic of low [$\alpha$/Fe]\citep[see also][]{marsakov18a}, further strengthening previous arguments in the literature that the most metal-rich RRL are drawn from the thin disc.    

Non-variable thin-disc stars are known to be younger on average than thick-disc ones from comparisons with isochrones in the H-R diagram \citep[e.g.,][]{haywood13}, which may also explain their lower $\alpha$/Fe ratios. The old ages ($\gtrsim 10$ Gyrs) of metal-poor RRL ([Fe/H] $\lesssim -0.8$) are well established by their presence in globular clusters, which have been accurately dated from their main-sequence turnoffs.  There is no such direct evidence for the more metal-rich RRL because the metal-rich globular clusters do not contain RRL on account of their very red horizontal branches \footnote{Unusually long-period RRL have been found in the metal-rich globular clusters NGC 6388 and 6441, which also contain blue HB stars, (see \citet{pritzl03}).  These RRL do not resemble in metal-rich RRL near the Sun.}.  It is likely that the RRL with thin-disc kinematics and [$\alpha$/Fe]$\sim 0$, are among the oldest of thin-disc stars, because with decreasing age it becomes more difficult for stars to lose sufficient mass to evolve into the instability strip at the luminosities of these stars.  Since only a small number of thin-disc RRL are found in a region occupied by a huge number of other thin disk stars, their production must be a rare event \citep{layden95b}.  These are interesting stars from the standpoints of their evolution and the chronology of the galactic discs, and worthy of further study \citep[see][]{bono97,marconi18}.
         
\subsection{The Halo RRL stars: relics of past mergers}

\subsubsection{Angular Momentum and Energy}

\begin{figure}
\includegraphics[angle=270,width=\columnwidth]{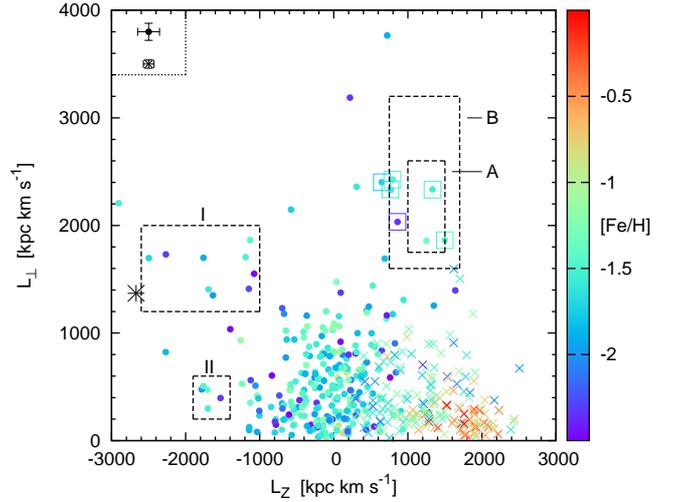}
\caption{The halo ($V_{pec} > 210$\,km \,s$^{-1}$) and `disc' RRL (solid circles and X's respectively) are plotted by angular momentum around the Z axis ($L_{Z}$) and out of the plane ($L_{\perp}$).  The RRL previously recognized as members of the ``Helmi-Stream'' have squares drawn around them.  The boxes labelled A and B were shown by \citet{koppelman19a} to contain many more stars that are likely members of this stream.  Groups I and II of RRL are defined by the labelled boxes.  The globular cluster NGC 3201 is plotted as a large asterisk. The error bars plotted in the upper left corner represent the mean errors of the RRL samples: halo top, `disc' bottom.}   
\label{fig:LzLper}
\end{figure}

\begin{figure}
\includegraphics[angle=270,width=\columnwidth]{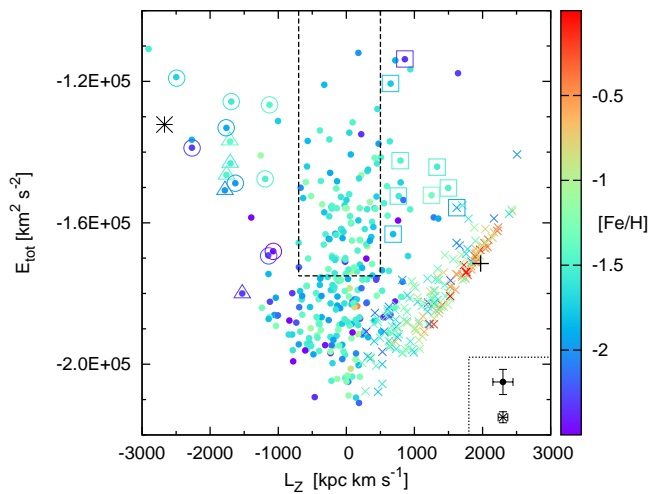}
\caption{The halo ($V_{pec} > 210$\,km \,s$^{-1}$) and `disc' RRL are plotted by $L_{Z}$ and total energy, using the same symbols as  Fig.~\ref{fig:LzLper}.  The members of the `Helmi-Stream' and Groups I and II have squares, circles and triangles drawn about them, respectively.  The globular cluster NGC 3201 is plotted as a large asterisk, and the LSR by a large cross.  The RRL in the Plume are contained within the rectangle defined by the dashed lines.  The error bars plotted in the lower right corner represent the mean errors of the samples: halo top, `disc' bottom.  The errors in $E_{tot}$ are based solely on the propagation of the errors in the components of the space velocities.}   
\label{fig:LzE}
\end{figure}

The plot of angular momentum out of the galactic plane, $L_{\perp}$, against angular momentum around the Z axis, $L_{Z}$, has been frequently used in discussions of halo substructures in the solar vicinity \citep[e.g.,][]{helmi99, refiorentin05, kepley07, kinman12, helmi18, koppelman18, li19}.  For our RRL sample, this diagram is shown in Fig.~\ref{fig:LzLper}, where the stars are defined as halo if $V_{pec} > 210$\,km \,s$^{-1}$ and otherwise as `disc', following \citet{helmi18}. This cut is designed to limit the contamination by disc stars in the halo sample. Some of the so-defined `disc RRL' have low values of $L_{Z}$ and/or fairly large values of $L_{\perp}$ (see Fig.~\ref{fig:LzLper}), and hence are more halo-like than disc-like.  These stars could be be part of the `in situ' halo that is found in simulations or the kinematically hot tail of the thick disc population, which may be one in the same. As one expects from their positions in the Toomre diagram, many of the metal-rich stars have values of $L_{Z}$ near that the LSR (1902 kpc~km~s$^{-1}$) and small values of $L_{\perp}$(see also Fig.~\ref{fig:LzE}). 

In the pioneering study by \citet{helmi99} of stars within 2.5 kpc of the Sun, a stellar stream, now called the `Helmi-Stream', was recognized as a clump of stars, including 3 RRL, in the $L_{Z} - L_{\perp}$ diagram.  Subsequent investigations \citep{refiorentin05, kepley07, kinman12} increased the number of RRL in the group to 7.  With the exception of AS Cnc, which \citet{kepley07} considered a dubious member and we concur on the basis of its large $L_{\perp}$ (3765 kpc km $\,s^{-1}$), these stars are identified in Fig.~\ref{fig:LzLper} along with two boxes labelled A and B.  According to \citet{koppelman19a}, Box A includes most of the stars making up the core members of the Helmi-Stream and Box B contains the remaining core members and many other stars that are likely members of the Stream. Fig.~\ref{fig:LzLper} shows that one additional star lies within box A and two more lie on the border of box B.  The RRL previously identified with the Helmi-Stream and these new ones are listed in Table~\ref{tab_groups}, where HS-A, HS-B, and HS-BB indicate if they are in box A, B, or borderline B, respectively.

We have put boxes, labelled I and II, around two other groups in Fig.~\ref{fig:LzLper} that have large retrograde values of $L_{Z}$. Their member stars are listed in Table~\ref{tab_groups}. Group I is approximately coincident in the $L_{Z} - L_{\perp}$ plane with the small group of stars discussed by \citet{kepley07} and \citet{kinman12}, which they pointed out may be part of a stellar stream.  Included in their group were two RRL, RV Cap and AT Vir, which are among the 9 RRL that we identify here as Group I. \citet{kinman12} noted that the globular cluster NGC 3201 lies not far from these stars in the $L_{Z} - L_{\perp}$ plane.  NGC3201 is plotted in Fig.~\ref{fig:LzLper} using the proper motion that \citet{baumgardt19} determined from \textit{Gaia} DR2 measurements and their tabulated values of $d_{\odot}$ and radial velocity.  With the more precise data from \textit{Gaia}, NGC 3201 is indeed close to the Group I RRL. Group II consists of 5 stars that are tightly clustered at large retrograde $L_{Z}$ and small $L_{\perp}$.  In the following diagrams, we identify the Helmi-Stream RRL, including the 3 new ones, and Groups I and II, so that the kinematic properties their member stars may be compared.

\begin{table}
\caption{Small Groups in the $L_Z - L_{\perp}$ plane.}
\label{tab_groups}
\begin{tabular}{cccc}
\hline
Name or ASAS & Group & Name or ASAS & Group \\
\hline
RZ~Cep              & HS-B  &  GV~Peg         & I \\
TT~Cnc              & HS-B  &  VX~Scl         & I \\ 
XZ~Cyg              & HS-A  &  MT~Tel         & I \\
CS~Eri              & HS-A  &  AT~Vir         & I \\
DI~Leo              & HS-BB &  002843-4400.4  & I \\
TT~Lyn              & HS-B  &  083443-3332.9  & I \\
AR~Ser              & HS-BB &  AQ~Cnc         & II \\
031119-2853.6       & HS-A  &  AR~Her         & II \\
221843-5652.4       & HS-BB &  RW~Hyi         & II \\
ZZ~And              & I &      TY~Pav         & II \\
RV~Cap              & I &      BP~Pav         & II \\
UW~Hor              & I &      ...            &    \\
\hline
\end{tabular}
\end{table}
     
In Fig.~\ref{fig:LzE} the RRL sample is plotted using $L_{Z}$ and $E_{tot}$. Co-moving groups of stars are expected to form clumps in Fig.~\ref{fig:LzE}, and the groups identified on the basis of Fig.~\ref{fig:LzLper} show signs of this.  Of the 9 RRL identified with the Helmi-Stream in Fig.~\ref{fig:LzLper}, 7 are clustered at $(L_{Z},E_{tot}) \sim (1200,-1.5 \times 10^5)$. The Helmi-Stream stars identified by \citet{koppelman19a} and the stars in their simulated stream occupy similar locations in the $L_{Z}-E_{tot}$ plane (see their figs. 2 and 15).  The 2 remaining Helmi-Stream RRL from Fig.~\ref{fig:LzLper} have larger values of $E_{tot}$ than the others, and their membership in the Stream is less secure.

Fig.~\ref{fig:LzE} shows that 7 of the 9 Group I RRL and 4 of the 5 Group II RRL are clustered in the $L_{Z} - E_{tot}$ plane as well as in Fig.~\ref{fig:LzLper}.  The evidence that they consitute streams is therefore about as strong  as the Helmi-Stream RRL.  Although the stars in Groups I and II overlap in $E_{tot}$, they are substantially different in $L_{\perp}$ because the Group II ones have small $V_{Z}$ velocities.  Several Group I stars are similar to NGC 3201 in $E_{tot}$ as well as $L_{Z}$ and $L_{\perp}$.  A subset of these are sufficiently close to NGC 3201 in $L_{Z}$ that they may be related. It is likely that Group I, and perhaps Group II as well, are part of the accretion event discussed very recently by \citet{myeong19}.  They propose that bulk of the high energy retrograde halo stars and several globular clusters including NGC 3201, originated in one large galaxy, called `Sequoia', which was disrupted as it merged with the MW \citep[see also][]{massari19}. In their discussion of multiple retrograde substructures in the local halo, \citet{koppelman19b} found evidence for `Sequoia', but also for an additional substructure of metal-poor stars at lower energy and $-800 \ge L_{Z} \ge -2000 $ kpc~km~s$^{-1}$, which they called `Thamnos'. There are several RRL in Fig.~\ref{fig:LzE} which occupy the same region, and are potentially members of this substructure.   

The most striking feature of Fig.~\ref{fig:LzE} is not, however, these moving groups.  It is the narrow range of $L_{Z}$ of the RRL with the largest energies.  This `Plume' of RRL at  $L_{Z} \sim 0$ runs from about the $E_{tot}$ of the LSR to higher values.  A rectangle has been drawn about it in Fig.~\ref{fig:LzE}, which is restricted in $L_{Z}$ to between +500 and -700 kpc~km~s$^{-1}$ to avoid confusion with the Helmi-Stream and the Group I stars.  There is no reason to suspect that the Plume does not extend to the bottom of Fig.~\ref{fig:LzE}, but at low $E_{tot}$ it may become more contaminated with stars of different origins.  We have therefore set the base of the rectangle at $E_{tot} = -1.75 \times 10^5\ \rm{km}^2 \ \rm{s}^{-2}$, where the Plume still appears to be a separate feature of the diagram.  

\citet{dinescu02} was the first to note this plume in the $L_{Z}-E_{tot}$ plane using data from the \citet{beers00} compilation of nearby metal-poor stars.  She suggested that it was the remains of the disrupted galaxy that contributed the globular cluster $\omega$ Cen to the MW.  This scenario was modelled by \citet{tsuchiya03} and also by \citet{brook03}, whose model exhibits a similar plume in the $L_{Z} - E_{tot}$ plane. The plume was also noted by \citet{morrison09} in their analysis of a sample of halo stars. The data from the \textit{Gaia} satellite show that the plume is a very prominent feature in the kinematics of local halo stars \citep{koppelman18,helmi18}, including many that are $\alpha$ poor \citep{haywood18, helmi18}.  The `Gaia-Sausage', which describes the anisotropy of the halo's velocity ellipsoid \citep{belokurov18b}, is basically the plume viewed in a different way.  There is consensus that the plume, called `Gaia-Enceladus' by \citet{helmi18}, and the `Gaia-Sausage' was produced by a major merger that provided many of stars of the galactic halo and also formed the thick disc by heating the pre-existing galactic disc \citep{belokurov18b, helmi18}.  This accreted galaxy probably also contributed a number of globular clusters to the halo \citep{helmi18,myeong18,massari19}. \citet{donlon19} have recently provided evidence that the Virgo Overdensity, the Virgo Stellar Stream, and other halo streams may be due one large merger, the `Virgo Radial Merger', and they suggest that this merger event may be the same as Gaia-Sausage/Enceladus.

\begin{figure}
\includegraphics[angle=0,width=\columnwidth]{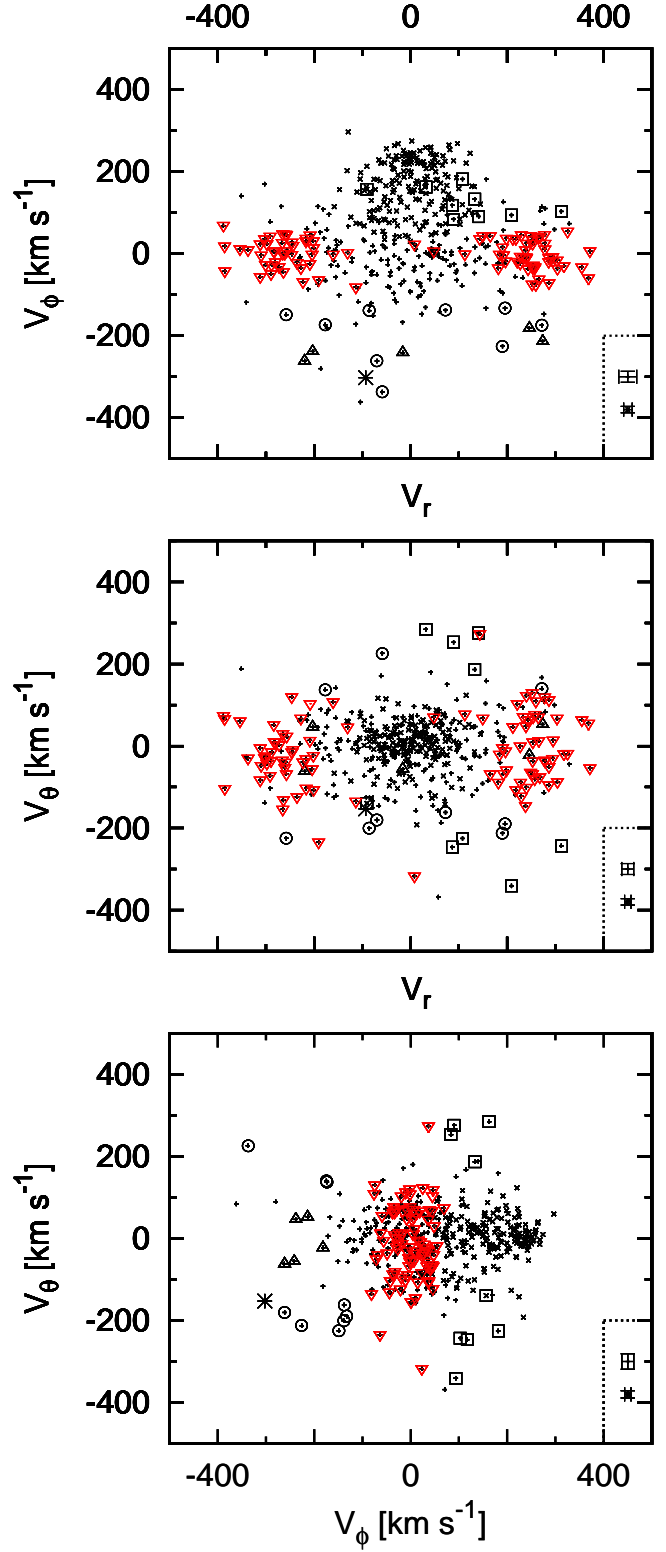}
\caption{The halo and `disc' RRL (+'s and x's respectively) are plotted using velocities in spherical coordinates.  Black squares, circles, and upward pointing triangles are drawn around the RRL in the Helmi-Stream, Group I, and Group II, respectively.  Downward pointing red triangles are drawn around the stars within the Plume rectangle in Fig.~\ref{fig:LzE}.  The globular cluster NGC 3201 is depicted by a large asterisk.  The mean error bars of the samples are plotted in the lower right corner of each diagram: halo top, `disc' bottom.}   
\label{fig:Velo}
\end{figure}

\subsubsection{Spherical Polar Coordinates and Anisotropy}

We have plotted the RRL using their velocities in spherical coordinates in Fig.~\ref{fig:Velo}.  $V_{r}$ is the velocity in the radial direction and $V_{\phi}$ and $V_{\theta}$ are the velocities in the azimuthal and polar directions, respectively. $V_{r}$ and $V_{\theta}$ were calculated from the cylindrical velocities $V_{R}$ and $V_{Z}$ using the following relations:

\begin{eqnarray}
V_{r} = V_{R}\frac{R}{r} + V_{Z}\frac{Z}{r},\\
V_{\theta} = V_{R}\frac{Z}{r} - V_{Z}\frac{R}{r},\\
r = (X^2 + Y^2 + Z^2)^{1/2}.
\end{eqnarray}

In Fig.~\ref{fig:Velo}, the Helmi-Stream and Groups I \& II RRL and the globular cluster NGC 3201 are identified so that their velocities may be compared.  The 93 RRL in the rectangle of Fig.~\ref{fig:LzE} have small values of $|V_{\phi}|$ as expected from their small $L_{Z}$.  These Plume stars make up the extremes in the $V_{r}$ distribution of the halo, which resembles the `sausage' distribution that \citet{belokurov18b} discovered in a large sample of main-sequence stars near the Sun.

Two of the RRL identified with the Plume in Fig.~\ref{fig:LzE}, SW Dor and ASAS 031948-3331.0, have much larger values of $|V_{\theta}|$ than the others.  Consequently, they are not considered Plume members in the following discussion.  These stars lie not far from the Helmi-Stream stars in Figs. 6, 7 and 8, and they could be members of that stream. 

\begin{figure}
\includegraphics[angle=270,width=\columnwidth]{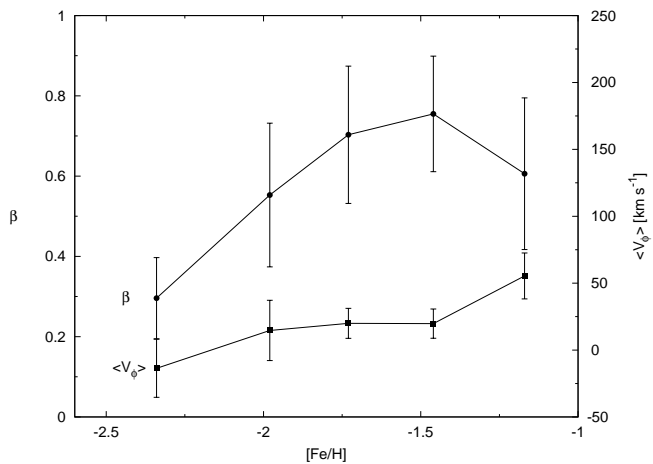}
\caption{For RRL with $|Z| \ge 1$ kpc, the anisotropy parameter $\beta$ and the mean azimuthal velocity <$V_{\phi}$> are plotted against [Fe/H].}
\label{fig:beta}
\end{figure}

Using their sample of main-sequence stars, \citep{belokurov18b} showed that the anisotropy parameter $\beta$\footnote{$\beta = 1-(\sigma_{\phi}^2 + \sigma_{\theta}^2)/2\sigma_{r}^2$. $\beta = -\infty, 0,\rm{\&} 1$ for purely circular, isotropic, and purely radial motions, respectively.} increases with increasing [Fe/H] and becomes quite large ($0.8-0.9$) in the range $-1.7 <$ [Fe/H] $< -1$. For the sample of RRL, plots of $\beta$ and mean azimuthal velocity against [Fe/H] are shown in Fig.~\ref{fig:beta}, where the sample has been limited to stars with $|Z| \ge 1.0$ kpc. A more extreme cut is necessary to eliminate the disc contamination, but this would limit too much the sample size (see the Toomre diagrams in Fig.~\ref{fig:toomre}). Inspection of the data reveals that the most metal-rich bin, which exhibits an uptern in <$V_{\phi}$> and a decline in $\beta$, is affected by disc-star contamination. The plot of $\beta$ in Fig.~\ref{fig:beta} is in qualitative agreement with the results of \citep{belokurov18b} in that the halo RRLS are moderately anisotropic at very low abundances and become much more so at [Fe/H]\ $> -1.7$.  Although $\beta$ is never $> 0.8$ in the RRLS sample, such extreme values are within the errors.  The plot of <$V_{\phi}$> shows that the halo is very slowly rotating over most of the [Fe/H] range, again in qualitative agreement with the main-sequence stars. 

The comparisons in this section and the previous ones show that the RRL sample has, at least in general terms, the kinematic signatures and kinematic-metallicity dependences as the other types of halo stars which have revealed the Gaia-Sausage/Enceladus merger.  This need not have been the case since the production of RRL in a stellar population is tied to its horizontal branch morphology, which is a complex function of age and chemical composition.

\subsection{The Properties of the Plume RRL}   

\begin{figure}
\includegraphics[angle=270,width=\columnwidth]{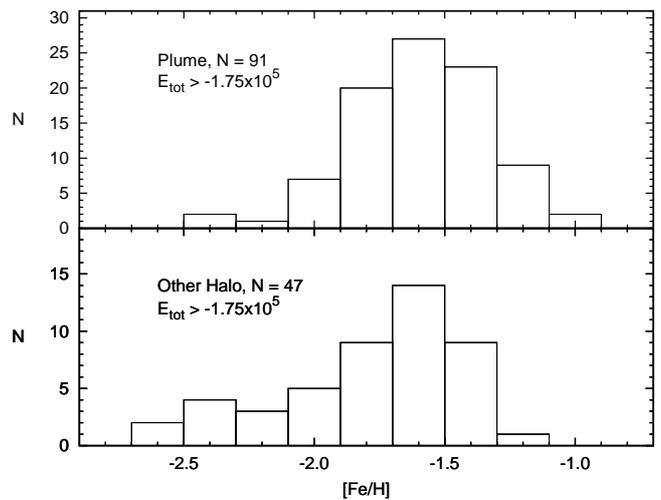}
\caption{The metallicity distribution functions of the RRL inside and outside the Plume.}
\label{fig:MDF}
\end{figure}

The metallicity distribution function (MDF) and the pulsational properties of the Plume RRL provide additional information on the Gaia-Sausage/Enceladus galaxy and how its very old stellar population compares with those of existing galaxies. In  Fig.~\ref{fig:MDF}, the MDF's of the Plume stars and halo stars of similar $E_{tot}$ but not part of the Plume (see Fig.~\ref{fig:LzE}) are compared.  The Plume and other halo stars have mean [Fe/H] values of $-1.58\pm0.03$ and $-1.77\pm0.05$, respectively, and Fig.~\ref{fig:MDF} shows that this difference is due to the Plume having a larger proportion of metal-rich stars and less of a metal-poor tail than the other halo stars. The Kolmogorov-Smirnov (K-S) 2-sided test rejects at the 10\% significance level the hypothesis that the two MDF's were drawn from the same parent. Since there is a tight correlation between galaxy metallicity and mass among dwarf galaxies \citep[e.g.,][]{kirby13}, these differences suggest that the Plume is the remains of a larger galaxy than the ones that contributed to the sample of other halo stars.

Halo stars divide into two groups by [$\alpha$/Fe], which also separates them by kinematics and age \citep{nissen10,schuster12}, in the sense that the low $\alpha$ stars are younger and have hotter kinematics.  Recent data from the APOGEE project has greatly increased the number of stars with measured abundances and reinforced the importance of the $\alpha$ abundances for interpreting the evolution of the MW  \citep[e.g.,][]{hayes18,fernandez18,mackereth19a}.  Using these data, \citet{helmi18}and \citet{haywood18} have shown that the stars making up Gaia-Sausage/Enceladus have, at the same [Fe/H], lower [$\alpha$/Fe] than thick disc stars, which suggest that it was slower in its metal enrichment than the MW disc.  We have examined the data that \citet{magurno18} compiled from high-resolution spectroscopic observations of RRL to see if the Plume stars exhibit a difference in [$\alpha$/Fe] compared to stars of similar [Fe/H].  The results were inconclusive because the Plume stars with measured $\alpha$ abundances have [Fe/H] < -1.4.  At these low values, there is at most a small offset between the Gaia-Sausage/Enceladus and the thick disk, and we could not rule out the possibility that the offset we observed with the data from \citet{magurno18}, which was in the predicted direction, was not due to the heterogeneity of the data sources.

\begin{figure}
\includegraphics[angle=270,width=\columnwidth]{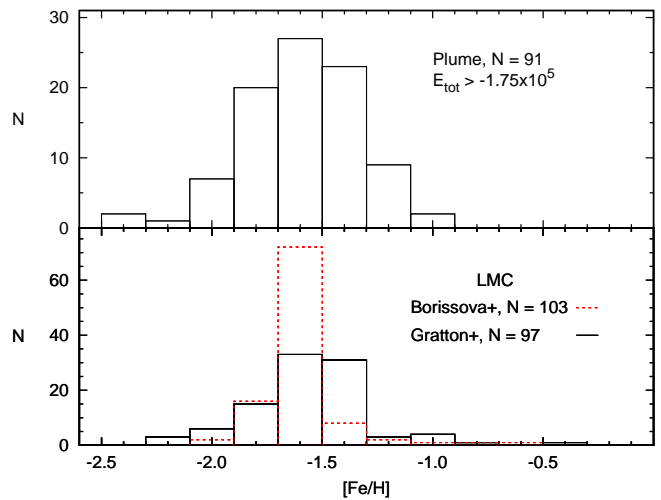}
\caption{The metallicity distribution functions of the Plume and two samples of RRL in the LMC \citep{gratton04,borissova06} are compared.}
\label{fig:MDF_LMC}
\end{figure}

\subsubsection{Comparison with the Large Magellanic Cloud}

A comparison of the Plume with existing MW satellite galaxies provides insight into the properties of the Gaia-Sausage/Enceladus galaxy, which has been estimated to be slightly more massive than the SMC \citep{helmi18} or more similiar to the LMC \citep{belokurov18b, haywood18, myeong18}.  The LMC appears to be the better comparison object because the <[Fe/H]> of the Plume (-1.58) is similar to the value obtained for the RRL in the LMC \citep[$\sim -1.6$,][]{gratton04, borissova06, haschke12a, wagner-kaiser13, skowron16}, and more metal rich than estimates for SMC RRL \citep[$\sim -1.8$,][]{kapakos12, skowron16}\footnote{[Fe/H] on the ZW scale which is identical to the one used here at these metallicities.}. Two investigations of LMC RRL have determined spectroscopically its MDF using a technique very similar in principle to the one employed here and in L94.  \citet{gratton04} measured 97 unblended RRL in two fields close to the bar of the LMC, and \citet{borissova06} measured 103 RRL in 8 fields in and near the bar using the same technique as \citet{gratton04}.  Their MDF's are shown in Fig.~\ref{fig:MDF_LMC}, which include the small offset \citep[-0.06,][]{gratton04} which places their [FeH] values on the ZW scale. The <[Fe/H]> values of the LMC MDF's are similar to each other, $-1.60\pm0.02$ and $-1.55\pm0.03$ for the \citet{borissova06} and \citet{gratton04} MDF's respectively, and to the Plume ($-1.58\pm0.03$).  However, the K-S test indicates at a very high significance (< 0.01) that the two LMC MDF's were not drawn from the same parent, because the \citet{borissova06} values are concentrated in a very narrow range.  Likewise the Plume has a significantly wider distribution than the \citet{borissova06} MDF. The Plume and the \citet{gratton04} MDF's are similar (see Fig.~\ref{fig:MDF_LMC}), and the K-S test indicates that there is no reason to reject the hypothesis that they were drawn from the same parent.

The MDF of the LMC has been also measured using either the technique of Fourier decomposition of their lightcurves \citep[e.g.,][]{haschke12a, skowron16} or from their periods and amplitudes \citep{wagner-kaiser13}.  These investigations, which have utilized thousands of RRL, have obtained <[Fe/H]> values near -1.6 on the ZW scale, which agree with the results obtained by \citet{gratton04,borissova06} and with the <[Fe/H]> of the Plume.  Because of systematic errors related to the Blazhko effect and in the tying of the Fourier results to the ZW scale \citep[see][]{haschke12b}, and because of the larger observational errors of these photometric methods, the shapes of their MDF's may be less reliable than the spectroscopic ones. Nonetheless, the Plume's similarity with the MDF measured by \citet{gratton04} and with several measurements of <[Fe/H]> are strong evidence that a LMC-size galaxy was its origin.     

\begin{figure}
\includegraphics[angle=270,width=\columnwidth]{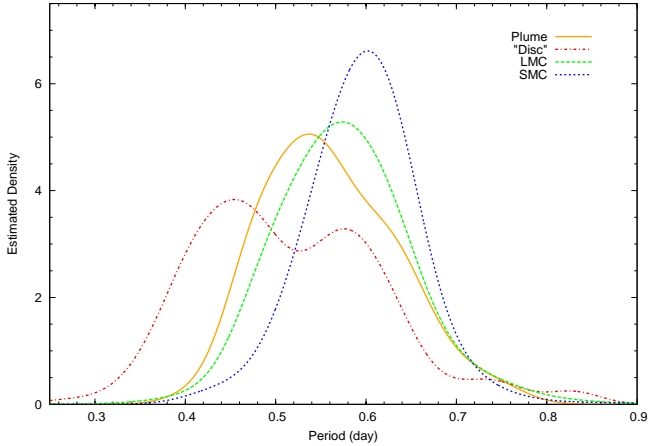}
\caption{For the type ab RRL, the estimated probability density of the periods of the Plume stars is compared with the ones for the `disc' (i.e., $V_{pec} < 210$ km$s^{-1}$) stars, the LMC, and the SMC. These densities were calculated using a Gaussian kernal with a standard deviation of 0.03 day.}
\label{fig:EstDen}
\end{figure}

The distribution in the periods of the type ab RRL is another diagnostic of the evolution of a stellar system.  As pointed out by \citet{zinn14}, the ab variables in the MW halo have a period distribution that extends to shorter periods than those of the dwarf spheroidal (dSph) galaxies of the lowest mass.  It is similar to the period distributions of the LMC and the Sgr dSph, which suggested that the accretion of a few large satellite galaxies, rather than many low-mass ones, was responsible for the build-up of the halo.  In Fig.~\ref{fig:EstDen}, the period distributions of the Plume stars are compared with those of the LMC \citep{soszynski09}, SMC \citep{soszynski10}, and the `disc' sample.  The Plume resembles most closely the LMC distribution, but it has a larger fraction of short period variables.  The two-tail K-S test rules out the null hypothesis that they have the same distribution at the 0.05 significance level.  The period distributions of the ab variables in globular clusters indicate there is a rough correlation between the periods of the shortest ab variables and cluster metallicity (see also below), in the sense that shortest periods are found in the most metal-rich clusters.  The period distributions suggest therefore that at the time when RRL formed ($\gtrsim 10$ Gyr. ago), the Plume (the Gaia-Sausage/Enceladus galaxy) was slightly more metal-rich than the LMC.  

\begin{figure}
\includegraphics[angle=270,width=\columnwidth]{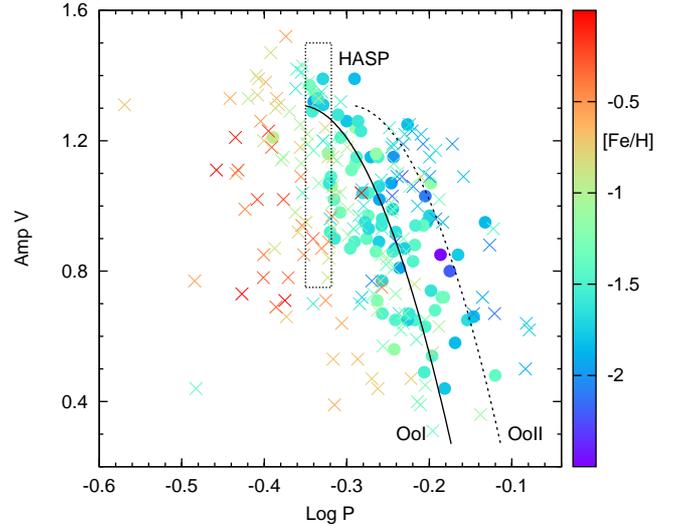}
\caption{The type ab RRL in the Plume (solid circles) and the `disc' (X's) are plotted in the Bailey diagram. Also plotted are the ridge lines of the Oosterhoff type I and type II globular clusters and, following \citet{fiorentino15}, a box delineating the HASP region, although all type-ab RRL with P $\le 0.48$ d and Amp $\ge 0.75$ are HASP stars by their definition.}
\label{fig:Pamp}
\end{figure}

As noted by \citet{fiorentino15}, the Bailey diagram, the plot of amplitude against period, is also a diagnostic of the evolution of a stellar system.  They draw particular attention to the region they call the High Amplitude Short Period (HASP) type ab RRL (see Fig.~\ref{fig:Pamp}, which they define as P\ $\le 0.48$ d and Amp\ $\ge 0.75$). The low-mass dSph galaxies and metal-poor globular clusters do not contain HASP variables, whereas HASP stars are found in the higher mass satellite galaxies, such as the Sgr dSph galaxy, the LMC and the SMC, and in relatively metal-rich globular clusters.  Because the MW halo has many HASP stars, \citet{fiorentino15} reached the same conclusion as \citet{zinn14}, that much of the halo population of the MW must have origined in massive satellites \citep[see also][]{fiorentino17}.  \citet{monelli17} have shown that the satellites of M31 behave in a similar way to the MW satellites in terms of the frequencies of HASP stars, and most importantly, that HASP populations are found only in massive satellities.  

In the Bailey diagram in Fig.~\ref{fig:Pamp}, we have plotted the Plume and `disc' samples of type-ab RRL.  It is customary to plot the locations of the ridge lines of the Oosterhoff type I and type II globular clusters in the Bailey diagram, and we show them here to point out that some variables in the samples are indistinguishable from OoI or OoII stars. Although there is substantial scatter in [Fe/H] at any particular location in Fig.~\ref{fig:Pamp}, there is a general trend in the sense that shorter period is correlated with larger [Fe/H].  While there are only 85 ab variables in the Plume sample, there are 12 in the HASP region of Bailey diagram, although 5 are arguably borderline. The resulting HASP fraction (8-14\%) is larger than than those of the LMC and SMC (6\% and 1\%, respectively, \citet{fiorentino15}). The Bailey diagram as well as MDF and the period distribution of the ab variables suggest that the Plume originated in a massive satellite. 

In Fig.~\ref{fig:MDF} we compared the Plume sample to other high energy halo stars, most of which belong to the Helmi Stream and Groups I \& II. There are 5 HASP stars in this sample, and each of these 3 groups contains one.  As noted above the retrograde Groups I \& II may be part of the `Sequoia' galaxy, which according to \citep{myeong19} was a massive satellite, although not as massive as the Gaia-Sausage/Enceladus. \citet{koppelman19a} have argued that the Helmi Stream probably originated in yet another massive satellite. The presence of some HASP RRL in these Groups and the Helmi Stream is not surprising. The other 2 HASP stars lie not far from the $L_Z$ boundaries of the Plume in Fig.~\ref{fig:LzE}, which were conservatively drawn to avoid contamination from the other groups.  The simulation of the Gaia-Sausage/Enceladus merger by \citet{helmi18} indicates that while its stars are concentrated in the $L_{Z}$ boundaries that define the Plume, some are scattered over a wider range.              

Figs.~\ref{fig:EstDen} and \ref{fig:Pamp} show that the `disc' sample, the RRL with $V_{pec} < 210$\,km$s^{-1}$, has many more short period ab variables than the Plume.  This sample is a mixture of thin- and thick-disc stars and halo stars with pro-grade rotations (see above).  The stars with the largest [Fe/H] and the shortest periods are members of the thin- or thick-discs and are therefore most likely native to the MW.  In order of increasing fractions of short period ab variables, the systems in Fig.~\ref{fig:EstDen} are: the SMC, the LMC, the Plume, and the MW.  This appears, therefore, to be the order in increasing [Fe/H], and probably also the ordering by galaxy mass.  
             
\section{Conclusions}

Improvements in the astrometric data (thanks to the \textit{Gaia} satellite) and an increase in the number of RRL stars with [Fe/H] measurements on a homogeneous scale has made it possible to examine more closely the kinematics and other properties of the RRL near the Sun.  These data confirm that the MW has a thin disc of RRL with near solar $\alpha$/Fe ratios \citep[see also][]{marsakov18b} in addition to the well-documented thick disc of RRL.  The oldest thin-disc stars of similar composition and kinematics to the RRL have ages that are typically a few Gyrs younger than the ages measured for metal-poor RRL via globular cluster dating \citep[e.g.,][]{mackereth19b}.  How a small fraction of these thin-disc stars were able to evolve into the instability strip and become the observed thin-disc RRL is unclear.       

Among the halo RRL, there are small groups that have similar locations in plots of $L_{\perp}$ and $E_{tot}$ against $L_{Z}$, which are signs that they originated in the same accretion events (the 'Helmi-Stream', `Sequoia', and 'Thamnos', see above).  The most obvious of these in our sample is the Helmi-Stream \citep[][and refs. therein]{koppelman19a}, which contains 9 and possibly 11 RRL.    

The most remarkable property of the local RRL population is, however, that the majority of high energy RRL are spread out in $E_{tot}$ while concentrated in a small range of $L_{Z}$ near zero, creating a `Plume' in the $L_{Z} - E_{tot}$ plane. This Plume is largely responsible for the marked change in radial anisotropy with increasing [Fe/H] in the RRL sample, which resembles the one \citet{belokurov18b} found in a sample of main-sequence stars.  In all respects, the kinematics of the Plume RRL closely resemble as the `Gaia-Sausage' \citep{belokurov18b} or `Gaia-Enceladus' \citep{helmi18}, which appear to have their origin in a large dwarf galaxy that merged with the MW \citep[see also][]{myeong18,kruijssen19,mackereth19a,vincenzo19,iorio19,fattahi19}. The MDF of the Plume RRL is skewed towards higher metallicity than the MDF of other high-energy RRL, many of which are members of either the prograde Helmi-Stream or retrograde groups that may be part of the `Sequoia' galaxy \citep{myeong19}. At least near the Sun, Gaia-Sausage/Enceladus is the major contributor of metal-rich RRL to the MW halo.  The <[Fe/H]> of the Plume and its MDF suggest that its stars originated in a large galaxy similar to the LMC. The period distribution of the type ab variables and the frequency of HASP stars indicate that this satellite galaxy may have even more massive than the LMC. The `disc' sample of RRL, which is largely native to the MW as opposed to the alien ones accreted from satellites, is more metal-rich and has higher frequency of short-period type ab than the Plume. At the time of RRL formation the MW had obtained a higher metallicity than the the Gaia-Sausage/Enceladus galaxy, whose later merger with the MW produced much of stellar halo we observe today.

\section*{Acknowledgements}

We acknowledge the geneous award of telescope time through Yale University's participation in the SMARTS consortium. We thank A. K. Vivas for discussions on the use of the spectrograph on the 1.5m CTIO Telescope.  We also thank the anonymous referee for comments and suggestions which improved this paper.
This work has made use of data from the European Space Agency (ESA) mission
{\it Gaia} (\url{https://www.cosmos.esa.int/gaia}), processed by the {\it Gaia}
Data Processing and Analysis Consortium (DPAC,
\url{https://www.cosmos.esa.int/web/gaia/dpac/consortium}). Funding for the DPAC
has been provided by national institutions, in particular the institutions
participating in the {\it Gaia} Multilateral Agreement.  We also made use of the SIMBAD database, operated at CDS, Strasbourg, France.




\bibliographystyle{mnras}











\bsp	
\label{lastpage}
\end{document}